\documentclass[aps,nofootinbib,groupedaddress,onecolumn,showkeys]{revtex4}
\usepackage{amssymb}
\usepackage{amsmath}
\usepackage{epsfig}

\input{tcilatex}

\begin{document}

\title{Bethe ansatz solution of discrete time stochastic processes with
fully parallel update. }
\author{A.M. Povolotsky$^{1,2}$ and J.F.F. Mendes$^1$}
\affiliation{$^1$Physics Department, University of Aveiro, Campus de Santiago - 3810-193,
Aveiro, Portugal}
\affiliation{$^2$Bogoliubov Laboratory of Theoretical Physics, Joint Institute for
Nuclear Research, Dubna 141980, Russia}

\begin{abstract}
We present the Bethe ansatz solution for the discrete time zero range and
asymmetric exclusion processes with fully parallel dynamics. The model
depends on two parameters: $p$, the probability of single particle hopping,
and $q$, the deformation parameter, which in the general case, $|q|<1$, is
responsible for long range interaction between particles. The particular
case $q=0$ corresponds to the Nagel-Schreckenberg traffic model with $v_{%
\mathrm{max}}=1$. As a result, we obtain the largest eigenvalue of the
equation for the generating function of the distance travelled by particles.
For the case $q=0$ the result is obtained for arbitrary size of the lattice
and number of particles. In the general case we study the model in the
scaling limit and obtain the universal form specific for the
Kardar-Parisi-Zhang universality class. We describe the phase transition
occurring in the limit $p\rightarrow 1$ when $q<0$.
\end{abstract}

\maketitle

\section{Introduction.}

One-dimensional models of stochastic processes attracted much attention last
decade. Being related to several natural phenomena like the interface growth
\cite{kpz}, the traffic flow\cite{ns}, and the self-organized criticality%
\cite{btw}, they admit an exact calculation of \ many physical quantities,
which can not be obtained with mean-field approach. Such models served as a
testing ground for the description of many interesting effects specific for
nonequilibrium systems like, boundary induced \ phase transitions \cite{Krug}%
,\cite{BoundPhase}, shock waves \cite{shocks} \ and condensation transition
\cite{evans}.

The most \ prominent in the physical community one-dimensional stochastic
model is the Asymmetric Simple Exclusion Process (ASEP). The interest in
this model was inspired by its Bethe ansatz solution \cite{dhar},\cite{gs},
which became the first \ direct exact calculation of the dynamical exponent
of Kardar-Parisi-Zhang equation \cite{kpz}. Since then plenty of exact
results were obtained in this direction using both the Bethe ansatz \cite%
{schutzrev} and the matrix product method \cite{derrida}. The advantage of
the former is the possibility of consideration of time dependent quantities
rather than only the stationary ones. Among the results obtained for the
ASEP with the help of the Bethe Ansatz there are the crossover scaling
functions for the Kardar-Parisi-Zhang universality class \cite{kim}, the
large deviation function of the Kardar-Parisi-Zhang-type interface \cite{dl},%
\cite{lk}, the time dependent correlation functions on the infinite \cite%
{schutz} and the periodic lattices \cite{priezz}.

Though the Bethe ansatz solvability \ opens rich opportunities for obtaining
exact results, it implies restrictive limitations for the dynamical rules,
such that the range of the Bethe ansatz solvable models is not too wide.
Besides the ASEP,\ the examples of the particle hopping models, which admit
the Bethe ansatz solution, are the asymmetric diffusion models \cite{sw,sw2}%
, generalized drop-push models \cite{srb,karim1,karim2}, and the asymmetric
avalanche process (ASAP) \cite{piph}. \ Most of the models of stochastic
processes solved by the Bethe ansatz are formulated in terms of continuous
time dynamics or random sequential update, which allows one to use the
analogy with the integrable quantum chains. The results for the \ processes
with discrete time parallel update are rare \cite{schutz1}. On the other
hand in conventional theory of quantum integrable systems the fundamental
role is played by the two dimensional vertex models, all the quantum spin
chains and continuous quantum models being their particular limiting cases
\cite{baxter}. In the theory of one-dimensional stochastic processes the
same role could be played by the models with discrete time parallel update,
i.e. so called stochastic cellular automata \cite{wolfram}. Having plenty of
real applications, the cellular automata give splendid opportunity for doing
large scale numerical simulations. Thus, finding the Bethe ansatz solutions
for discrete time models with fully parallel update would be of interest.
The lack of such solutions owes probably to more complicated structure of
the stationary state of models with the fully parallel update, which makes
the application of the Bethe ansatz more subtle.

Recently \ the Bethe ansatz was applied to solve the continuous time
zero-range process (ZRP) \cite{spitzer} with the nonuniform stationary state
\cite{pov}. \ The solution was based on the Bethe ansatz weighted with the
stationary weights of corresponding configurations. The simple structure of
the stationary state\ of the ZRP allowed finding the one-parametric family
of the hopping rates, which provides the Bethe ansatz integrability of the
model. In the present article we use the same trick to solve much more
general model of\ the ZRP with fully parallel update, which is defined by
two-parametric family of hopping probabilities. As particular limiting cases
of the parameters the model includes q-boson asymmetric diffusion model\cite%
{sw2}, which in turn includes the drop-push \cite{srb} and phase model \cite%
{bik}, and the ASAP \cite{piph}. Simple mapping allows one to consider also
the ASEP-like model that obeys the exclusion rule. In general it includes
the long range interaction between particles, which makes the hopping
probabilities dependent on the length of queue of particles next to the
hopping one. In the simplest case, when the latter interaction is switched
off, the model reduces to the Nagel-Schreckenberg traffic model \cite{ns}
with the maximal velocity $v_{\max }=1$. In the article we find the
eigenfunctions and eigenvalues\ of the equation for the generating function
of the distance travelled by particles and obtain the Bethe equations for
both ASEP and ZRP cases. We analyze the solution of the Bethe equations
corresponding to the largest eigenvalue \ of the equation for the generating
function of the distance travelled by particles. The analysis shows that a
particular limit of parameters of the model reveals the second order phase
transition. When the density of particles reaches the critical value, the
new phase emerges, which changes the dependence of the average flow of
particles on the density of particles. The phase transition turns out to be
intimately related to the intermittent-to-continuous flow transition in the
ASAP, and the jamming transition in the traffic models. We analyze this
transition in detail.

The article is organized as follows. In the section II we formulate the
model and discuss the basic results of the article. In the section III we
give the Bethe ansatz solution of the equation for the generating function
of the moments of the distance travelled by particles. In the section IV we
study the long time behavior of this equation and obtain its largest
eigenvalue. The special form of the Bethe equations in the particular case,
which corresponds to the Nagel-Schreckenberg model with $v_{\max }=1$,
allows one to proceed with the calculations for arbitrary size of the
lattice and number of particles. In the other cases we obtain the results in
the scaling limit. In the section V we discuss the particular limit of the
model, which exhibits the phase transition. In the section VI we discuss the
calculation of the stationary correlation functions with the partition
function formalism. A short summary is given in the section VII.

\section{Model definition and main results.}

\subsection{ Zero range process.}

The system under consideration can be most naturally formulated in terms of
the discrete time ZRP. Let us consider the one-dimensional periodic lattice
consisting of $N$ sites with $M$ particles on it. Every site can hold any
number of particles. A particle from an occupied site jumps to the next site
forward with probability $p(n)$, which depends only on the occupation number
$n$ of a site of departure. The system evolves step by step in the discrete
time $t$, all sites being updated simultaneously at every step. \ Thus, as a
result of the update the configuration $C$, defined by the set of occupation
numbers%
\begin{equation}
C=\left\{ n_{1},\ldots ,n_{N}\right\} ,  \label{C}
\end{equation}%
changes as follows
\begin{equation}
\left\{ n_{1},\ldots ,n_{N}\right\} \rightarrow \left\{
n_{1}-k_{2}+k_{1},\ldots ,n_{N}-k_{1}+k_{N}\right\} .
\end{equation}%
Here the variable $k_{i}$ , taking values $1$ or $0$, \ denotes the number
of particles arriving at the site $i$. \ According to these dynamical rules,
all $k_{i}$-s associated to the same time step are independent random
variables with the distribution, which depends on the occupation number of
the site $i-1$%
\begin{equation}
P\left( k_{i}=1\right) =p\left( n_{i-1}\right) ,P\left( k_{i}=0\right)
=1-p\left( n_{i-1}\right) .
\end{equation}%
The probability $P_{t}\left( C\right) $ for the system to be in a
configuration $C$ at time step $t$ obeys the Markov equation%
\begin{equation}
P_{t+1}(C)=\sum_{\left\{ C^{\prime }\right\} }T\left( C,C^{\prime }\right)
P_{t}\left( C^{\prime }\right) ,  \label{Master1}
\end{equation}%
where $T\left( C,C^{\prime }\right) $ is the probability of the transition
from $C^{\prime }$ to $\ C$. This equation is known to have a unique
stationary solution \cite{evans}, which belongs to the class of the so
called product measures, i.e. the probability of a configuration is given by
the product of one-site factors
\begin{equation}
P_{st}\left( n_{1},\ldots ,n_{N}\right) =\frac{1}{Z\left( N,M\right) }%
\prod_{i=1}^{N}f\left( n_{i}\right) ,  \label{P_st}
\end{equation}%
where the one-site factor $f\left( n\right) $ is defined as follows%
\begin{eqnarray}
f\left( 0\right) &=&1-p\left( 1\right) ,  \notag \\
f\left( n\right) &=&\frac{1-p\left( 1\right) }{1-p\left( n\right) }%
\prod_{i=1}^{n}\frac{1-p\left( i\right) }{p\left( i\right) },\,\,n>0
\label{f(n)}
\end{eqnarray}%
and $Z\left( N,M\right) $ is the normalization constant. The properties of
this stationary measure have been intensively investigated particularly due
to the possibility of condensation transition in such systems \cite{evans}.
The question we are going to address below is: which dynamics leading to
such a stationary measure admits the exact solution?

\subsection{Associated exclusion process.}

\begin{figure}[tbp]
\unitlength=1mm \makebox(110,30)[cc] {\psfig{file=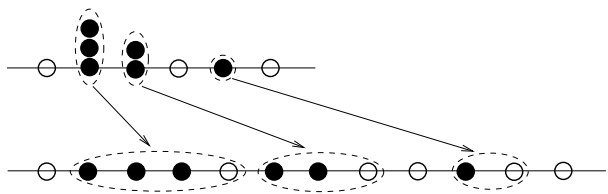,width=80mm}}
\caption{ZRP-ASEP mapping. Each site occupied with $n$ particles is mapped
to the sequence of $n$ sites occupied each by one particle with one empty
site ahead.}
\label{ASEP-ZRP}
\end{figure}

A simple mapping allows one to convert ZRP into ASEP. Given a ZRP
configuration, we represent every site occupied with $n$ particles as a
sequence of $n$ sites occupied each with one particle plus one empty site
ahead (see Fig. \ref{ASEP-ZRP}).
Thus, we add $M$ extra sites to the lattice, so that its \ size becomes $%
L=M+N.$ Obviously the exclusion constraint is imposed now. The evolution
started with such a configuration is completely defined by the above ZRP
dynamics. Specifically, any particle, which belongs to a cluster of $n$
particles and has empty site ahead (i.e. the first particle of the cluster),
\ jumps forward with the probability $p\left( n\right) $ or stays with the
probability $\left( 1-p\left( n\right) \right) $. Thus, we obtain the
generalized ASEP with the asymmetric long range interaction, which we refer
to as the ASEP associated to given ZRP. Note that the above ZRP-ASEP
relation applied on the periodic lattice establishes correspondence between
the sequences of configurations rather than between the configurations
themselves. This fact can be illustrated with a simple example. Consider the
following sequence of steps of the ZRP evolution. At the first step a
particle jumps from an occupied site $i$ to the next site $i+1$, at the
second from $i+1$ to $i+2$ and so on until a particle comes to the site $i$
from $i-1$ restoring the initial configuration. Reconstruction of the
corresponding sequence of the ASEP configurations implies that when a
particle joins a cluster of particles from the behind, the front particle of
the same cluster takes the next step, which leads to the shift of the
cluster one step backward. As a result we finally come to the configuration
translated as a whole one step backward with respect to the initial one.
This difference does not allow the ZRP and the ASEP to be considered as one
model. There are, however, some quantities, which are not sensitive to
translations\footnote{%
The exact meaning of the term "insensitive to translations" will be
clarified in the section III.}, and therefore are identical for two models.
For example the stationary measure of the associated ASEP is given by the
expression (\ref{P_st}), though $f\left( n\right) $ corresponds now to a
cluster of $n$ particles with one empty site ahead. Apparently, the
probability of any configuration does not change under the translation of
the configuration as a whole. Below we consider the generating function of
the total distance travelled by particles, $Y_{t}$, in the infinite time
limit, $lim_{t\rightarrow \infty }\ln \left\langle e^{\gamma
Y_{t}}\right\rangle /t$ ,which is also the same for the ZRP and the ASEP
associated to it. The only thing we should keep in mind in such cases is
that in ZRP the numbers $N$ and $M$ denote the lattice size and the number
of particles respectively, while in the associated ASEP $N$ \ is the number
of holes, $M$ is the number of particles, and the size of the lattice is $%
L=N+M$. Of course, time- or spatially- dependent quantities in general
require separate consideration for each model. In this article we first
consider ZRP dynamics and then point out the difference\ of solution for the
of associated ASEP.

In the end of the discussion of the ZRP-ASEP correspondence we should
mention that another version of the ASEP can be obtained from the above one
by the particle-hole transformation. Then the hopping probabilities will
depend on the length of headway in front of the particle like for example
the hopping rates in the bus route model \cite{evans_BRM}. Though the latter
formulation seems more naturally related to the real traffic, we will use
the former one to keep the direct connection with the ZRP. Of course the
results for both versions can be easily related to each other.

\subsection{The results.}

To formulate basic result of the article we introduce the following
generating function
\begin{equation*}
F_{t}\left( C\right) =\sum_{Y=0}^{\infty }e^{\gamma Y}P_{t}\left( C,Y\right)
,
\end{equation*}%
where $P_{t}\left( C,Y\right) $ is the joint probability for the system to
be in configuration $C$ at time $t$, the total distance travelled by
particles $Y_{t}$ being equal to $Y$. By definition \ \ the generating
function, $F_{t}\left( C\right) $, coincides with the probability $%
P_{t}\left( C\right) $ of the configuration $C$ in the particular case $%
\gamma =0$ . The function $F_{t}(C)$ obeys the evolution equation similar to
(\ref{Master1})
\begin{equation}
F_{t+1}(C)=\sum_{\left\{ C^{\prime }\right\} }e^{\gamma \mathcal{N}\left(
C,C^{\prime }\right) }T\left( C,C^{\prime }\right) F_{t}\left( C^{\prime
}\right) .  \label{Master F_t(C)}
\end{equation}%
The term $e^{\gamma \mathcal{N}\left( C,C^{\prime }\right) }$ accounts the
increase of the total distance $Y_{t}$, $\mathcal{N}\left( C,C^{\prime
}\right) $ being the number of particles, which make a step during the
transition from $C^{\prime }$ to $C$. \ \ Below we show that the
eigenfunctions of this equation, satisfying the following eigenfunction
problem%
\begin{equation}
\Lambda \left( \gamma \right) F_{_{\Lambda }}(C)=\sum_{\left\{ C^{\prime
}\right\} }e^{\gamma \mathcal{N}\left( C,C^{\prime }\right) }T\left(
C,C^{\prime }\right) F_{\Lambda }\left( C^{\prime }\right) ,
\label{eigenproblem}
\end{equation}%
have the form of the Bethe ansatz weighted with the weights of stationary
configurations provided that the hopping probabilities has the following form%
\begin{equation}
p\left( n\right) =p\times \left[ n\right] _{q},\quad n=1,2,3,\ldots ,
\label{p(n)}
\end{equation}%
where $\left[ n\right] _{q}$ are the so called q-numbers
\begin{equation}
\left[ n\right] _{q}=\frac{1-q^{n}}{1-q}.  \label{q-numbers}
\end{equation}%
Thus, as usual, the Bethe ansatz to be applicable, the infinite in general
set of toppling probabilities should be reduced to two parametric family,
the parameters being $p$ and $q$. Formally, there are no any further
limitations on the parameters . However, once we want \ $p\left( n\right) $
to be probabilities, they should be less than or equal to one and
nonnegative. It is obviously correct for any $n$ if
\begin{equation}
0\leq p\leq 1-q\quad \mathrm{and\quad }\left\vert q\right\vert \leq 1.
\label{q<1}
\end{equation}%
If $q>1$, and an integer $n^{\ast }>1$ exists, such that $\ 1/\left[ n^{\ast
}+1\right] _{q}<p<1/\left[ n^{\ast }\right] _{q}$, \ we can still formulate
the model with a finite number of particles $M<n^{\ast }$. The only way to
consider an arbitrary number of particles with $q>1$ is to consider the
limit $p\rightarrow +0$. \ Particularly, if we put $p=\delta \tau ,$ where $%
\delta \tau $ is infinitesimally small, the model turns into the\textit{\ }%
continuous time version of ZRP (or associated ASEP) known also as q-boson
asymmetric diffusion model \cite{sw}. In this model the Poisson rate $u(n)$
of the hopping of a particle from a site is given by $u(n)=[n]_{q}$, $q$
taking on values in the range$\ q\in \left( -1,\infty \right) $. Such
continuous time limit seems to be the only possibility for $q$ to exceed $1$%
. This model was studied \ in \cite{sw},\cite{pov}. \ The case $(p>1-q,$ $%
\left\vert q\right\vert \leq 1)$ also implies $p\left( n\right) >1$ for some
finite $n>n^{\ast }$ and thus do not allow consideration of arbitrary $M$,
which is of practical interest. Thus, below we concentrate on the domain (%
\ref{q<1}).

The other particular limits of the model under consideration, which
reproduce the models studied before are as follows:

-- \textit{The Nagel-Schreckenberg traffic model with }$v_{\max }=1$
corresponds to the ASEP with $q=0$, when the probabilities $p\left( n\right)
$ are independent of $n$, $p\left( n\right) \equiv p$. This case will be
considered separately because, due to the special factorized form of the
Bethe equations, it can be treated for arbitrary finite $N$ and $M$.

-- \textit{The asymmetric avalanche process} corresponds to the limit%
\begin{equation}
(1-p)\rightarrow 0.  \label{ASAP rate}
\end{equation}%
Originally the ASAP has been formulated as follows. In a stable state, $M$
particles are located on a ring of $N$ sites, each site being occupied at
most by one particle. At any moment of continuous time each particle can
jump one step forward with Poisson rate $1$. If a site $i$ contains more
than one particle $n_{i}>1$, it becomes unstable, and must relax immediately
by spilling forward either $n$ particles with probability $\mu _{n}$ or $n-1$
particles with probability $1-\mu _{n}$. The relaxation stops when all sites
become stable again with $n_{i}\leq 1$ for any $i$. The period of subsequent
relaxation events is called avalanche. The avalanche is implied to be
infinitely fast with respect to the continuous time. The toppling
probabilities, which ensure the exact solvability of the model can also be
written in terms of $q-$numbers%
\begin{equation}
\mu _{n}=1-[n]_{q},\quad -1<q\leq 0,  \label{mu_n}
\end{equation}%
Below we show that the ASAP \ can be obtained as the continuous time limit
of the ZRP observed from the moving reference frame, one step of the
discrete time $t$ being associated with the \ infinitesimal time interval $%
(1-p)$.

Let us now turn to the physical results we can extract from the Bethe ansatz
solution. To this end, we note that the large time behavior of the
generating function of the distance travelled by particles $\left\langle
e^{\gamma Y_{t}}\right\rangle $ is defined by the largest eigenvalue $%
\Lambda _{0}\left( \gamma \right) $ of the equation (\ref{eigenproblem})
\begin{equation*}
\left\langle e^{\gamma Y_{t}}\right\rangle =\sum_{\left\{ C\right\}
}F_{t}\left( C\right) \sim \left( \Lambda _{0}\left( \gamma \right) \right)
^{t},\quad t\rightarrow \infty .
\end{equation*}%
Respectively, the logarithm of the largest eigenvalue $\Lambda _{0}\left(
\gamma \right) $ is the generating function of the cumulants of $Y_{t}$ in
the infinite time limit
\begin{equation*}
\lim_{t\rightarrow \infty }\frac{\left\langle Y_{t}^{n}\right\rangle _{c}}{t}%
=\left. \frac{\partial ^{n}\ln \Lambda _{0}\left( \gamma \right) }{\partial
\gamma ^{n}}\right\vert _{\gamma =0}.
\end{equation*}%
Here the angle brackets denote the cumulants rather than moments of the
particle current distribution, which is emphasized with the subscript $c$.
Below we give the results for $\Lambda _{0}\left( \gamma \right) $ given in
terms of two parameters, one of which is $q$ introduced above, and the other
is $\lambda $ defined as follows
\begin{equation}
\lambda =\frac{p}{1-p-q},  \label{lambda}
\end{equation}%
which captures all the dependence on the above $p$.

* \textit{For }$q=0$\textit{, which corresponds to the Nagel-Schreckenberg
traffic model with unit maximal velocity, }$v_{\max }=1$\textit{,} we obtain%
\begin{eqnarray}
\ln \Lambda _{0}\left( \gamma \right) &=&-\lambda \sum_{n=1}^{\infty }\frac{%
B^{n}}{n}\binom{Ln-2}{Nn-1}\left. _{2}F_{1}\right. \left(
\begin{array}{c}
1-Mn,1-nN \\
2-nL%
\end{array}%
;-\lambda \right) ,  \label{Lambda, q=0} \\
\gamma &=&-\frac{1}{M}\sum_{n=1}^{\infty }\frac{B^{n}}{n}\binom{Ln-1}{Nn}%
\left. _{2}F_{1}\right. \left(
\begin{array}{c}
-Mn,-nN \\
1-nL%
\end{array}%
;-\lambda \right) ,  \label{gamma, q=0}
\end{eqnarray}%
where $\Lambda _{0}\left( \gamma \right) $ is defined parametrically, both $%
\Lambda _{0}\left( \gamma \right) $ and $\gamma $ being power series in the
variable $B$, $\binom{a}{b}=\frac{a!}{b!\left( a-b\right) !}$ is a binomial
coefficient, {\small \ }$\left. _{2}F_{1}\right. \left(
\begin{array}{c}
a,b \\
c%
\end{array}%
;x\right) $ is the Gauss hypergeometric function, and $L=N+M$.

* \textit{For an arbitrary }$q$\textit{\ we obtain the results in the
scaling limit} $N\rightarrow \infty ,$ $M\rightarrow \infty ,$ $M/N=\rho
=const,$ $\gamma N^{3/2}=const$ \ \ \ \
\begin{equation}
\ln \Lambda _{0}(\gamma )=N\phi \gamma +k_{1}N^{-3/2}G(k_{2}N^{3/2}\gamma ),
\label{ln Lamda_0}
\end{equation}%
where $G(x)$ is a universal scaling function defined parametrically, both $%
G(x)$ and $x$ being the functions of the same parameter $\varkappa $
\begin{eqnarray}
G(x) &=&-\sum_{s=1}^{\infty }\frac{\varkappa ^{s}}{s^{5/2}}  \label{G(x)} \\
x &=&-\sum_{s=1}^{\infty }\frac{\varkappa ^{s}}{s^{3/2}}.  \label{x(C)}
\end{eqnarray}%
This function appeared before in the studies of the ASEP \cite{dl}, ASAP
\cite{pph2} and polymers in random media \cite{DB} believed to belong to the
KPZ universality class. \ The \ parameters $\phi ,k_{1},k_{2}$ are the model
dependent constants, which depend on two parameters $\lambda $ and $q$ and
the density of particles $\rho $. \ An explicit form of these constants will
be derived in the following sections.

The results in the limit $p\rightarrow 1,q<0$ are of special interest. Below
we show that in this limit the model exhibits a phase transition, which is
closely related to the intermittent-continuous flow transition in the ASAP
and the jamming transition in traffic models. When the density of particles
is less than critical density%
\begin{equation}
\rho _{c}=1/(1-q),  \label{rho_c}
\end{equation}%
almost all particles synchronously jump forward with rare exceptions, which
happen with probability of order of $(1-p)$. Thus the average flow of
particles is equal to the density of particles with small correction of
order of $(1-p)$.%
\begin{equation}
\phi \left( \rho <\rho _{c}\right) \simeq \rho
\end{equation}%
However, when the density approaches $\rho _{c}$ the average flow $\phi $
stops growing.%
\begin{equation}
\phi \left( \rho \geq \rho _{c}\right) \simeq \rho _{c}
\end{equation}%
It turns out that a fraction of particles gets stuck at a small fraction of
sites instead of being involved into the particle flow. As a result the
density of the particles involved into the flow is kept equal to $\rho _{c}$%
. \ This phenomenon can be treated as a phase separation. The new phase
consisting of immobile particles emerges at the critical point. Any increase
of the density above the critical point leads to the growth of this phase,
while the density of mobile particles remains unchanged. In \ terms of the
associated ASEP or the traffic models one can describe the situation as an
emergence of \ a small number of traffic jams with average length diverging,
when $(1-p)$ decreases. Such behaviour leads to singularities in the
constants $\phi ,k_{1},k_{2}$, Eq.(\ref{ln Lamda_0}), which define the large
deviations of the particle current from the average. Below we analyze them
in detail.

We also study the stationary measure of the model with the canonical
partition function formalism. The calculation of the stationary correlation
functions are discussed. We observe the relations between the Bethe ansatz
solution in the thermodynamic limit and the thermodynamic properties of the
stationary state. We make a conjecture that the relations found between the
parameters characterizing the large deviations of the particle current and
the parameters of the stationary state hold for the general ZRP belonging to
the KPZ universality class. Finally we analyze the change of the behaviour
of the occupation number distribution at the critical point in the limit $%
p\rightarrow 1$. In addition the evaluation of $Z\left( N,M\right) $ reveals
interesting relation of the model with the theory of $q$-series \cite{aar},
being based on the use of the most famous relation of this theory, $q$%
-binomial theorem.

\section{ Bethe ansatz solution.}

Let us consider the equation for the generating function $F_{t}\left(
C\right) $ for \ the ZRP,
\begin{eqnarray}
F_{t+1}\left( n_{1},\ldots ,n_{N}\right)  &=&\sum_{\left\{ k_{i}\right\}
}\prod_{i=1}^{N}\left( e^{\gamma }p\left( n_{i}-k_{i}+k_{i+1}\right) \right)
^{k_{i+1}}  \notag \\
&&\times \left( 1-p\left( n_{i}-k_{i}+k_{i+1}\right) \right) ^{1-k_{i+1}}
\label{master} \\
&&\times F_{t}\left( n_{1}-k_{1}+k_{2},\ldots ,n_{N}-k_{N}+k_{1}\right) ,
\notag
\end{eqnarray}%
where we imply periodic boundary conditions $N+1\equiv 1$ . The summation is
taken over all possible values of $k_{1},\ldots ,k_{N}$, which are the
numbers of particles arriving at sites $1,\ldots ,$ $N$ $\ $respectively.
They take on the values of either $0$ or $1$ if $n_{i}\neq 0$ and of only $0$
otherwise.

Before using the Bethe ansatz for the solution of this equation we should
make the following remark. Most of the models studied by the coordinate
Bethe ansatz have a common property. That is, a system evolves to the
stationary state, where all the particle configurations occur with the same
probability. This property can be easily understood from the structure of
the Bethe eigenfunction. Indeed, the stationary state is given by the
groundstate of the evolution operator, which is the eigenfunction with zero
eigenvalue and momentum. Such Bethe function does not depend on particle
configuration at all and results in the equiprobable ensemble. Apparently
the ZRP under consideration is not the case like this. The way how to reduce
the problem to the one with uniform groundstate was proposed in \cite{pov}.
The main idea is to look for the solution in the form,%
\begin{equation}
F_{t}\left( C\right) =P_{st}\left( C\right) F_{t}^{0}\left( C\right) .
\label{F^0def}
\end{equation}%
Here $P_{st}\left( C\right) $ is the stationary probability defined in (\ref%
{P_st}). It is not difficult to check that $F_{t}^{0}\left( C\right) $
satisfy the following equation%
\begin{eqnarray}
F_{t+1}^{0}\left( n_{1},\ldots ,n_{N}\right) &=&\sum_{\left\{ k_{i}\right\}
}\prod_{i=1}^{N}\left( e^{\gamma }p\left( n_{i}\right) \right)
^{k_{i}}\left( 1-p\left( n_{i}\right) \right) ^{1-k_{i}}  \notag \\
&&\times F_{t}^{0}\left( n_{1}-k_{1}+k_{2},\ldots ,n_{N}-k_{N}+k_{1}\right) .
\label{master2}
\end{eqnarray}%
This form of the equation has two important advantages. The first is that
the coefficient before $F_{t}^{0}\left( C^{\prime }\right) $ under the sum
is the product of one site factors. The second is that if we formally define
$p(0)=0,$ the values of $k_{i}$-s will take on values of $0$ and $1$ with no
respect to the value of $n_{i}$. To proceed further we should go to a
different way of representation of system configurations. Let us define the
configuration by the coordinates of particles written in non-decreasing order%
\begin{equation}
C=\left\{ x_{1},\ldots ,x_{M}\right\} ,\quad x_{1}\leq x_{2}\leq \cdots \leq
x_{M}.  \label{domain}
\end{equation}%
Obviously this is nothing but the change of notations, both representations
being completely equivalent. Below we refer to them as to the occupation
number representation and the coordinate representation respectively. We are
going to show that the eigenfunctions of the equation (\ref{master2})
written in the coordinate representation can be found in form of the Bethe
ansatz. We first consider the cases $M=1$ and $M=2$ subsequently
generalizing them for an arbitrary number of particles.

\subsection{The case $M=1$.}

The eigenfunction problem for the master equation for one particle is simply
the one for the asymmetric random walk in discrete time.%
\begin{equation}
\Lambda \left( \gamma \right) F^{0}\left( x\right) =e^{\gamma }pF^{0}\left(
x-1\right) +\left( 1-p\right) F^{0}\left( x\right)   \label{Master,M=1}
\end{equation}%
The eigenfunction is to be looked for in the form%
\begin{equation}
F^{0}\left( x\right) =z^{-x},  \label{ansatz,M=1}
\end{equation}%
where $z$ is some complex number. Substituting (\ref{ansatz,M=1}) to (\ref%
{Master,M=1}) we obtain the expression of the eigenvalue
\begin{equation*}
\Lambda \left( \gamma \right) =e^{\gamma }pz+\left( 1-p\right) .
\end{equation*}%
The periodic boundary conditions
\begin{equation*}
F^{0}\left( x+N\right) =F^{0}\left( x\right)
\end{equation*}%
imply the limitations on the parameter $z$%
\begin{equation*}
z^{N}=1.
\end{equation*}

\subsection{The case $M=2$.}

In this case we consider two particles with coordinates $x_{1}$ and $x_{2}$.
If $x_{1}\neq x_{2}$ the equation is that for the asymmetric random walk for
two non-interacting particles
\begin{eqnarray}
\Lambda \left( \gamma \right) F^{0}\left( x_{1},x_{2}\right) &=&\left(
e^{\gamma }p\right) ^{2}F^{0}\left( x_{1}-1,x_{2}-1\right)  \notag \\
&&+e^{\gamma }p\left( 1-p\right) \left( F^{0}\left( x_{1}-1,x_{2}\right)
+F^{0}\left( x_{1},x_{2}-1\right) \right)  \notag \\
&&+\left( 1-p\right) ^{2}F^{0}\left( x_{1},x_{2}\right) ,  \label{master,M=2}
\end{eqnarray}%
while the case $x_{1}=x_{2}=x$ should be treated separately.%
\begin{equation}
\Lambda \left( \gamma \right) F^{0}\left( x,x\right) =e^{\gamma }p\left(
2\right) F^{0}\left( x-1,x\right) +\left( 1-p\left( 2\right) \right)
F^{0}\left( x,x\right)  \label{master,M=2,x1=x2}
\end{equation}%
The general strategy of the Bethe ansatz solution is as follows. We want to
limit ourselves by the only functional form of the equation for $F^{0}\left(
C\right) $ of the form (\ref{master,M=2}). To get rid of the additional
constraints imposed by interaction, like that in (\ref{master,M=2,x1=x2}),
we recall that the function $F^{0}\left( x_{1},\ldots ,x_{M}\right) $, is
defined in the domain$\quad x_{1}\leq x_{2}\leq \cdots \leq x_{M}$. If we
formally set $x_{1}=x_{2}=x$ in the equation (\ref{master,M=2}), one of the
terms in r.h.s., $F^{0}\left( x,x-1\right) $, will be outside of this
domain. We could redefine it in such a way, that the Eq.(\ref%
{master,M=2,x1=x2}) would be satisfied. This leads us to the following
constraint.%
\begin{equation}
aF^{0}\left( x,x\right) +bF^{0}\left( x,x-1\right) +cF^{0}\left(
x-1,x\right) +dF^{0}\left( x-1,x-1\right) =0,  \label{constraint,M=2}
\end{equation}%
where%
\begin{eqnarray*}
a &=&\left( \left( 1-p\right) ^{2}-\left( 1-p\left( 2\right) \right) \right)
, \\
b &=&e^{\gamma }p\left( 1-p\right) , \\
c &=&e^{\gamma }\left( p\left( 1-p\right) -p\left( 2\right) \right) , \\
d &=&\left( e^{\gamma }p\right) ^{2}
\end{eqnarray*}%
The solution of the equation (\ref{master,M=2}) can be looked for in the
form of the Bethe ansatz,%
\begin{equation}
F^{0}\left( x_{1},x_{2}\right)
=A_{1,2}z_{1}^{-x_{1}}z_{2}^{-x_{2}}+A_{2,1}z_{1}^{-x_{2}}z_{2}^{-x_{1}}.
\label{ansatz, M=2}
\end{equation}%
Substituting it into Eq.(\ref{master,M=2}) we obtain the expression for the
eigenvalue%
\begin{equation*}
\Lambda \left( \gamma \right) =\left( e^{\gamma }pz_{1}+\left( 1-p\right)
\right) \left( e^{\gamma }pz_{2}+\left( 1-p\right) \right) ,
\end{equation*}%
while the constraint (\ref{constraint,M=2}) results in the relation between
the amplitudes $A_{1,2}$ and $A_{2,1}$.%
\begin{equation*}
\frac{A_{1,2}}{A_{2,1}}=-\frac{a+bz_{1}+cz_{2}+dz_{1}z_{2}}{%
a+bz_{2}+cz_{1}+dz_{1}z_{2}}.
\end{equation*}%
The last step, which makes the scheme self-consistent, is to impose periodic
boundary conditions%
\begin{equation*}
F^{0}\left( x_{1},x_{2}\right) =F^{0}\left( x_{2},x_{1}+N\right) ,
\end{equation*}%
which lead us to the system of two algebraic equations. The first one is
\begin{equation*}
z_{1}^{-N}=-\frac{a+bz_{1}+cz_{2}+dz_{1}z_{2}}{a+bz_{2}+cz_{1}+dz_{1}z_{2}}
\end{equation*}%
and the second is obtained by the change $z_{1}\longleftrightarrow z_{2}$.

\subsection{The case of arbitrary $M$.}

For an arbitrary number of particles we will follow the same strategy. We
consider the equation for noninteracting particles, which jump forward with
the probability $p$
\begin{eqnarray}
\Lambda \left( \gamma \right) F^{0}\left( x_{1},\ldots ,x_{M}\right)
&=&\sum_{\left\{ k_{i}\right\} }\prod_{i=1}^{M}\left( e^{\gamma }p\right)
^{k_{i}}\left( 1-p\right) ^{1-k_{i}}  \notag \\
&&\times F^{0}\left( x_{1}-k_{1},\ldots ,x_{M}-k_{M}\right) .
\label{master3free}
\end{eqnarray}%
Here, all the numbers $k_{i}$-s run over values $1$ and $0$. \ Our aim is to
redefine the terms beyond the physical domain, (\ref{domain}), in terms of
ones within this domain in order to satisfy the equation with ZRP
interaction (\ref{master2}) written in the coordinate representation. This
redefinition results in many constraints on the terms $F^{0}\left( \ldots
\right) $. As it will be seen below, the Bethe ansatz to be applicable all
these constraints should be reducible to the only one,%
\begin{eqnarray}
0 &=&aF^{0}\left( \ldots ,x,x,\ldots \right) +bF^{0}\left( \ldots
,x,x-1,\ldots \right)  \notag \\
&&+cF^{0}\left( \ldots ,x-1,x\ldots \right) +dF^{0}\left( \ldots
,x-1,x-1,\ldots \right) ,  \label{constraint}
\end{eqnarray}%
which has been found for the two particle case. To this end, we first recall
that the r.h.s. of Eq.(\ref{master2}) is the sum of the terms $F^{0}\left(
\ldots \right) $ corresponding to configurations, from which the system can
come to the configuration in l.h.s. with coefficients factorized into the
product of one-site terms. Therefore, the processes corresponding to a
particle arriving at a site can be considered for each site separately.
Consider, for instance, Eq.(\ref{master2}), where the argument of the
function in l.h.s. is a configuration with a site $x$ occupied by $n$
particles.%
\begin{align}
\Lambda \left( \gamma \right) F^{0}\left( \ldots ,\left( x\right)
^{n},\ldots \right) & =\sum\nolimits_{\left\{ k_{i}\right\} }^{\prime
}\prod\nolimits_{i\neq x}^{\prime }\left( e^{\gamma }p\left( n_{i}\right)
\right) ^{k_{i}}\left( 1-p\left( n_{i}\right) \right) ^{1-k_{i}}  \notag \\
& \times \left[ e^{\gamma }p\left( n\right) F^{0}\left( \ldots ,\left(
x-1\right) ,\left( x\right) ^{n-1},\ldots \right) \right.  \label{master3} \\
& \quad \quad \quad \left. +\left( 1-p\left( n\right) \right) F^{0}\left(
\ldots ,\left( x\right) ^{n},\ldots \right) \right]  \notag
\end{align}%
The terms of r.h.s. can be grouped in pairs shown in square brackets, which
correspond to the processes, in which a particle comes or does not come to
the site $x.$ Here, $\left( x\right) ^{n}$ denotes $n$ successive arguments
equal to $x$, i.e. the site $x$ is occupied by $n$ particles. The primed
summation and product run over all sites apart of $x$. Obviously the
coefficients in the equations for noninteracting particles can be factorized
in similar way, such that the term corresponding to the site $x$ looks as
follows%
\begin{eqnarray}
&&\sum_{k_{1}=0}^{1}\cdots \sum_{k_{n}=0}^{1}\left( e^{\gamma }p\right)
^{\sum_{i=1}^{n}k_{i}}\left( 1-p\right) ^{n-\sum_{i=1}^{n}k_{i}}
\label{nonint} \\
&&\;\;\;\;\;\;\;\;\;\;\;\;\;\;\;\;\;\;\;\;\times F^{0}\left( \ldots
,x-k_{1},\ldots ,x-k_{n},\ldots ,\right) .  \notag
\end{eqnarray}%
We want this term to be equal to the one in the square brackets of Eq.(\ref%
{master3}). If $n=2$ the form of the term in square brackets is similar to
the r.h.s of Eq.(\ref{master,M=2,x1=x2}) and can be treated (i.e. reduced to
the noninteracting form (\ref{nonint})) using the constraint (\ref%
{constraint}). The equality for general $n$ can be proved by induction.
Suppose it is valid for $n-1$. Then we can perform the summations in Eq.(\ref%
{nonint}) over $k_{i}$-s for $i=2,\ldots ,n$, resulting in
\begin{eqnarray}
&&\left( 1-p\right) \left[ e^{\gamma }p\left( n-1\right) F^{0}\left( \ldots
,x,\left( x-1\right) ,\left( x\right) ^{n-2},\ldots \right) \right.
\label{interact} \\
&&\left. +\left( 1-p\left( n-1\right) \right) F^{0}\left( \ldots ,\left(
x\right) ^{n},\ldots \right) \right]  \notag \\
&&+e^{\gamma }p\left[ e^{\gamma }p\left( n-1\right) F^{0}\left( \ldots
,\left( x-1\right) ^{2},\left( x\right) ^{n-2},\ldots \right) \right.  \notag
\\
&&\left. +\left( 1-p\left( n-1\right) \right) F^{0}\left( \ldots ,\left(
x-1\right) ,\left( x\right) ^{n-1},\ldots \right) \right] .  \notag
\end{eqnarray}%
The first summand contains the term $F^{0}\left( \ldots ,x,\left( x-1\right)
,\left( x\right) ^{n-2},\ldots \right) $, which, being beyond the physical
region (\ref{domain}), can be expressed in terms of ones inside the physical
region by using the constraint (\ref{constraint}). \ Equating the
coefficients of the terms $F^{0}\left( \ldots \right) $ with the same
arguments of Eqs.(\ref{nonint}) and (\ref{interact}) we obtain the relation
between $p\left( n\right) $ and $p\left( n-1\right) $%
\begin{equation}
p\left( n\right) =p+qp\left( n-1\right) ,  \label{recrel}
\end{equation}%
where we introduce the notation $q$ defined as follows
\begin{equation}
p\left( 2\right) \equiv p\times \left( 1+q\right) .
\end{equation}%
The recurrent relation (\ref{recrel}) can be solved in terms of $q-$numbers
as was claimed in Eqs.(\ref{p(n)},\ref{q-numbers}).

Thus, we have shown that in the physical domain (\ref{domain}) the solution
of the equation for ZRP coincides with the solution of the equation for
non-interacting particles if the latter satisfies the constraint (\ref%
{constraint}). Then we can use the Bethe ansatz
\begin{equation}
F^{0}(x_{1},\ldots ,x_{M})=\sum_{\{\sigma _{1},\ldots ,\sigma
_{M}\}}A_{\sigma _{1},\ldots ,\sigma _{M}}\prod_{i=1}^{M}z_{\sigma
_{i}}^{-x_{i}}  \label{Bethe ansatz}
\end{equation}%
for the eigenfunction of the free equation. Here $z_{1},\ldots ,z_{M}$ are
complex numbers, the summation is taken over all $p!$ permutations $\{\sigma
_{1},\dots ,\sigma _{M}\}$ of the numbers $(1,\ldots ,M)$. Substituting the
Bethe ansatz (\ref{Bethe ansatz}) into the Eq.(\ref{master3free}), we obtain
the expression for the eigenvalue%
\begin{equation}
\Lambda \left( \gamma \right) =\prod_{i=1}^{M}\left( e^{\gamma
}pz_{i}+\left( 1-p\right) \right) .  \label{Lambda(gamma)}
\end{equation}%
Substituting it to the constraint (\ref{constraint}) we obtain the relation
between pairs of the amplitudes $A_{\sigma _{1},\ldots ,\sigma _{M}}$, which
differ from each other only in two indices permuted
\begin{equation}
\frac{A_{\cdots j,i\cdots }}{A_{\cdots i,j\cdots }}=S_{i,j}\equiv -\frac{%
a+bz_{j}+cz_{i}+dz_{i}z_{j}}{a+bz_{i}+cz_{j}+dz_{i}zj}.
\end{equation}%
With the help of this relation one can obtain all the amplitudes $A_{\sigma
_{1},\ldots ,\sigma _{M}}$ $\ $ in terms of only one, say $A_{1,\cdots ,M}$,
by successive permutations of indices, which results in multiplication by
the factors $S_{ij}$. For example, for three particle case we have%
\begin{eqnarray}
A_{2,1,3}
&=&S_{1,2}A_{1,2,3},\;\;A_{2,3,1}=S_{1,3}S_{1,2}A_{1,2,3},\;%
\;A_{3,1,2}=S_{2,3}S_{1,3}A_{1,2,3}  \notag \\
A_{1,3,2} &=&S_{2,3}A_{1,2,3},\;\;A_{3,2,1}=S_{2,3}S_{1,3}S_{1,2}A_{1,2,3}.
\end{eqnarray}%
In general, the procedure to be uniquely defined, it should be consistent
with the structure of the permutation group. Specifically, if $\widehat{%
\mathbf{\sigma }}_{i}$ is an elementary transposition that permutes the
numbers at $i$-th and $(i+1)$-th positions, it satisfies the following
relations.
\begin{eqnarray}
\widehat{\mathbf{\sigma }}_{i}\widehat{\mathbf{\sigma }}_{i+1}\widehat{%
\mathbf{\sigma }}_{i} &=&\widehat{\mathbf{\sigma }}_{i+1}\widehat{\mathbf{%
\sigma }}_{i}\widehat{\mathbf{\sigma }}_{i+1}  \notag \\
\widehat{\mathbf{\sigma }}_{i}^{2} &=&1  \label{permutation group}
\end{eqnarray}%
If the numbers at positions $i$, $\left( i+1\right) $, and $\left(
i+2\right) $ are $j,k,$ and $l$ respectively, two relations of Eq.(\ref%
{permutation group}) are reduced to
\begin{eqnarray}
S_{j,k}S_{j,l}S_{k,l} &=&S_{k,l}S_{j,l}S_{j,k},  \notag \\
S_{j,k}S_{k,j} &=&1,
\end{eqnarray}%
which are apparently true.

The last step we need to do is to impose the periodic boundary conditions.%
\begin{equation}
F^{0}\left( x_{1,},\ldots ,x_{M}\right) =F^{0}(x_{2},,\ldots ,x_{M},x_{1}+N)
\end{equation}%
They are equivalent to the following relations for the amplitudes.%
\begin{equation*}
A_{\sigma _{1},\ldots ,\sigma _{M}}=A_{\sigma _{2},\ldots ,\sigma
_{M},\sigma _{1}}z_{\sigma _{1}}^{-N}
\end{equation*}%
Consistency of this relation with Eq.(\ref{constraint}) results in the
system of $M$ algebraic Bethe equations,%
\begin{equation}
z_{i}^{-N}=\left( -\right) ^{M-1}\prod_{i=1}^{M}\frac{%
a+bz_{i}+cz_{j}+dz_{i}z_{j}}{a+bz_{j}+cz_{i}+dz_{i}z_{j}}.  \label{BAE ZRP}
\end{equation}%
We should emphasize that the crucial step for the Bethe ansatz solvability
is the proof that all many-particle interactions can be reduced to the
two-particle constraint, Eq.(\ref{constraint}). Existence of any other
constraints on the eigenfunctions would result in new equations for
parameters $z_{i},$ which would make the resulting system of algebraic
equations overdetermined.

\subsection{Bethe ansatz for the ASEP.}

As it was noted above, the Bethe ansatz solution for the ASEP is quite
similar to that for the ZRP and can be done in parallel. Indeed, if we
define the function $F_{t}^{0}\left( C\right) $ in the same way as it was
defined for ZRP, the equations for it can be obtained from those for ZRP by
the variable change
\begin{equation}
\left\{ x_{1,}x_{2},\ldots ,x_{M}\right\} \rightarrow \left\{
x_{1},x_{2}+1,\ldots ,x_{M}+M-1\right\} ,
\end{equation}%
which corresponds to the ZRP-ASEP transformation described above. Thus, the
free equation does not change its form, with the only difference that the
physical domain for the ASEP implies every site to be occupied at most by
one particle.
\begin{equation}
x_{1}<x_{2}<\cdots <x_{M}.
\end{equation}%
Therefore the Bethe ansatz (\ref{Bethe ansatz}) substituted to the free
equation results in the same form of the eigenvalue (\ref{Lambda(gamma)}).
The two particle constraint, which is used now to redefine the\ nonphysical
terms containing the pair $\left( x,x\right) $, have the following form
\begin{equation}
0=aF^{0}\left( x,x+1\right) +bF^{0}\left( x,x\right) +cF^{0}\left(
x-1,x+1\right) +dF^{0}\left( x-1,x\right) .
\end{equation}%
If we insert here the Bethe ansatz (\ref{Bethe ansatz}), we obtain the
following relation for the amplitudes%
\begin{equation}
\frac{A_{1,2}}{A_{2,1}}=-\frac{az_{1}^{-1}+b+cz_{2}z_{1}^{-1}+dz_{2}}{%
az_{2}^{-1}+b+cz_{1}z_{2}^{-1}+dz_{1}}.
\end{equation}%
All the other arguments, which extend the problem to the general $M-$%
particle, case are completely the same as those for the ZRP. Finally, the
periodic boundary conditions lead to the Bethe equations%
\begin{equation}
z_{i}^{-L}=\left( -\right) ^{M-1}\prod_{j=1}^{M}\frac{%
az_{i}^{-1}+b+cz_{j}z_{i}^{-1}+dz_{j}}{az_{j}^{-1}+b+cz_{i}z_{j}^{-1}+dz_{i}}%
.  \label{BAE ASEP}
\end{equation}%
Here we recall that the lattice size is equal now to $L=N+M$ rather than $N$%
. This equations can be rewritten in the following form%
\begin{equation}
z_{i}^{-N}=T\times \left( -\right) ^{M-1}\prod_{j=1}^{M}\frac{%
a+bz_{i}+cz_{j}+dz_{i}z_{j}}{a+bz_{j}+cz_{i}+dz_{i}z_{j}}.
\end{equation}%
where we introduce the notation
\begin{equation}
T=\prod_{j=1}^{M}z_{j}.
\end{equation}%
This term is the only difference between the Bethe equations for the ZRP and
the ASEP. Taking a product of all the equations we obtain,
\begin{equation}
T^{L}=1.
\end{equation}%
The term $T$ has a very simple physical meaning. This is the factor that
corresponding eigenfunction multiplies by under the unit translation. Thus,
the eigenfunction that corresponds to the set $\{z_{i}\}$ satisfying a
relation $T=1$ is invariant with respect to any translations for both ZRP
and ASEP. This is what we meant mentioning the translational invariance
above. Apparently the solutions of the Bethe equations, which satisfy the
relation $T=1$ coincide for the ZRP and the ASEP and so do any quantities
constructed with them. Below we consider such quantity and, therefore, do
not make a distinction between the ZRP and the ASEP.

\section{The largest eigenvalue.}

To proceed further, we introduce new variables $y_{i}$ defined as follows
\begin{equation}
z_{i}=e^{-\gamma }\frac{1-y_{i}}{1+\lambda y_{i}},
\end{equation}%
where $\lambda $ is defined in Eq.(\ref{lambda}). In these variables the
Bethe equations (\ref{BAE ZRP}) and the eigenvalue (\ref{Lambda(gamma)})
simplify to the following form%
\begin{eqnarray}
\left( \frac{1-y_{i}}{1+\lambda y_{i}}\right) ^{-N}e^{\gamma N} &=&\left(
-\right) ^{M-1}\prod_{j=1}^{M}\frac{y_{i}-qy_{j}}{y_{j}-qy_{i}},
\label{BAEy} \\
\Lambda \left( \gamma \right)  &=&\prod_{i=1}^{M}\frac{1+\lambda qy_{i}}{%
1+\lambda y_{i}}  \label{Lambda(gamma)y}
\end{eqnarray}%
Let us now consider the eigenstate, which corresponds to the maximal
eigenvalue $\Lambda _{0}\left( \gamma \right) $. Note that in the limit $%
\gamma \rightarrow 0$ the equation (\ref{Master F_t(C)}) for $F_{t}(C)$
turns into the Markov equation for the probability (\ref{Master1}). The
largest eigenvalue of the Markov equation is equal to 1. The corresponding
eigenstate is the stationary state (\ref{P_st}). It then follows from the
definition, that $F_{t}^{0}(C)$ in this limit becomes constant, i.e. it is
the same for all configurations, and the corresponding solution of the Bethe
equations is $z_{i}=1$ or $y_{i}=0$. If $\gamma $ deviates from $0$ the
analyticity and no-crossing of the eigenvalue is guaranteed by
Perron-Frobenius theorem. By continuity we also conclude that the parameter $%
T$ defined above is equal to 1 for arbitrary $\gamma $,
\begin{equation}
\prod_{j=1}^{M}e^{-\gamma }\frac{1-y_{j}}{1+\lambda y_{j}}=1.
\label{prod z=1}
\end{equation}

\subsection{The case $q=0$.}

In the case $q=0$, the form of Eq.(\ref{BAEy}) allows one to obtain the
largest eigenvalue for arbitrary $M$ and $N$ \cite{dl}. If we define the
parameter,
\begin{equation}
B=\left( -\right) ^{M-1}e^{\gamma N}\prod_{j=1}^{M}y_{j},
\end{equation}%
the solution of the Bethe equations, which corresponds to the largest
eigenvalue, will be given by $M$ roots of the polynomial%
\begin{equation}
P(y)=\left( 1+\lambda y\right) ^{N}B-\left( 1-y\right) ^{N}y^{M},
\end{equation}%
which approach zero when $B$ tends to zero. Then the sum of a function $%
f\left( x\right) $ analytic in some vicinity of zero over the Bethe roots,
can be calculated as the following integral over the contour closed around
the roots:
\begin{equation}
\sum_{j=1}^{M}f\left( y_{j}\right) =\oint \frac{dy}{2\pi i}\frac{P^{\prime
}\left( y\right) }{P\left( y\right) }f\left( y\right) .
\end{equation}%
Particularly, after the integration by parts the expression for $\ln \Lambda
_{0}\left( \gamma \right) $ has the following form:%
\begin{equation}
\ln \Lambda _{0}\left( \gamma \right) =\oint \frac{dy}{2\pi i}\frac{\lambda
}{1+\lambda y}\ln \left( 1-\frac{B\left( 1+\lambda y\right) ^{N}}{\left(
1-y\right) ^{N}y^{M}}\right) .  \label{Labda(B)}
\end{equation}%
At the same time the expression for $\gamma $ as a function of $B$ can be
obtained by taking a logarithm of Eq.(\ref{prod z=1})%
\begin{equation}
\gamma =\frac{1}{M}\oint \frac{dy}{2\pi i}\left( \frac{1}{1-y}+\frac{\lambda
}{1+\lambda y}\right) \ln \left( 1-\frac{B\left( 1+\lambda y\right) ^{N}}{%
\left( 1-y\right) ^{N}y^{M}}\right) .  \label{gamma(B)}
\end{equation}%
To evaluate it we make a series expansion of the logarithms in powers of $B$
and integrate the resulting series term by term. The procedure is valid
until the resulting series are convergent. The resulting series, obtained
after some algebra by using standard identities for hypergeometric
functions, are given in Eqs.(\ref{Lambda, q=0},\ref{gamma, q=0}). The large
time asymptotics of the cumulants of the distance travelled by particles $%
\lim_{t\rightarrow \infty }\left\langle Y_{t}^{n}\right\rangle _{c}/t$ can
be obtained by eliminating the parameter $B$ between two series. For
instance, the exact value of the average flow of particles is of the
following form:%
\begin{equation}
\phi =\frac{1}{N}\lim_{t\rightarrow \infty }\frac{\left\langle
Y_{t}\right\rangle _{c}}{t}=\frac{M\lambda }{L-1}\frac{\left.
_{2}F_{1}\right. \left(
\begin{array}{c}
1-M,1-N \\
2-L%
\end{array}%
;-\lambda \right) }{\left. _{2}F_{1}\right. \left(
\begin{array}{c}
-M,-N \\
1-L%
\end{array}%
;-\lambda \right) }.  \label{vexact}
\end{equation}%
In the thermodynamic limit $N\rightarrow \infty ,M\rightarrow \infty ,M/L=c$
\ the average velocity of particles $\overline{v}=N\phi /M$ \ saturates to
the formula obtained in \cite{ns},
\begin{equation}
\bar{v}=\frac{N}{M}\phi =\frac{1-\sqrt{1-4pc(1-c)}}{2c}.
\end{equation}%
It is straightforward to obtain the other cumulants up to any arbitrary
order. It is not difficult also to study the asymptotic behavior of the
series in the large $N$ \ limit. To this end one can evaluate the saddle
point asymptotics of the integrals instead of their exact values or use the
asymptotic formulas for the binomial coefficients and the hypergeometric
functions. The thermodynamic limit however allows significant simplification
already at the stage of writing of the Bethe equations, which makes possible
to consider more general case of arbitrary value of $q$.

\subsection{The case of arbitrary $q$.}

The technique used in this section was first developed to study the
non-factorizable Bethe equations for the partially ASEP \cite{kim,lk} and
then applied to the study of the ASAP \cite{pph2}. Since the analysis is
quite similar, we outline only the main points of the solution. Technical
details can be found in the original papers.

The scheme consists of the following steps. We assume that in the
thermodynamic limit, $N\rightarrow \infty ,M\rightarrow \infty ,M/N=\rho $,
the roots of the Bethe equations (\ref{BAEy}) are distributed in the complex
plain along some continuous contour $\Gamma $ with the analytical density $%
R(x)$, such that the sum of values of an analytic function $\mathrm{f}(x)$
at the roots is given by
\begin{equation}
\sum_{i=1}^{p}\mathrm{f}(x_{i})=N\int_{\Gamma }\mathrm{f}(x)R(x)dx.
\end{equation}%
After taking the logarithm and replacing the sum by the integral, the system
of equations (\ref{BAEy}) can be reformulated in terms of a single integral
equation for the density.
\begin{equation}
-\ln \left( \frac{1-x}{1+\lambda x}\right) +\gamma -\int_{\Gamma }\mathrm{%
\ln }\left( \frac{x-qy}{y-qx}\right) R(y)dy=i\pi \rho -2\pi
i\int_{x_{0}}^{x}R\left( x\right)   \label{int_eq}
\end{equation}%
The r.h.s. comes from the choice of branches of the logarithm, which
specifies particular solution corresponding to the largest eigenvalue. The
choice coincides with that appeared in the solution of the asymmetric
six-vertex model at the conical point \cite{bs} and later of the ASEP \cite%
{kim}. It can be also justified by considering the limit $q=0$, where the
locus of the roots can be found explicitly. The point $x_{0}$ corresponding
to the starting point of roots counting can be any point of the contour. \
The integral equation should be solved for a particular form of the contour,
which is not known \textit{a priory}, and being first assumed should be
self-consistently checked after the solution has been obtained. In practice,
however, an analytic solution is possible in the very limited number of
cases. Fortunately, we can proceed by analogy with the solution \ of the
asymmetric six-vertex model at the conical point \cite{bs} choosing the
contour closed around zero. The solution of the equation (\ref{int_eq})
yields the density%
\begin{equation}
R^{\left( 0\right) }(x)=\frac{1}{2\pi ix}\left( \rho -g_{q,\lambda
}(x)\right) ,  \label{R_0(x)}
\end{equation}%
where the function $g_{q,\lambda }(x)$ is defined as follows%
\begin{equation}
g_{q,\lambda }(x)=\sum_{n=1}^{\infty }\frac{1-\left( -\lambda \right) ^{n}}{%
1-q^{n}}x^{n}.  \label{g(x)}
\end{equation}%
This case corresponds to $\gamma =0$ and hence $\Lambda _{0}(\gamma )=1$.
Since $R^{\left( 0\right) }(x)$ is analytic in the ring $0<|x|<\lambda $,
the integration of it along any contour closed in this ring does not depend
on its form. Therefore, to fix the form of the contour additional
constraints are necessary. Such a constraint was found by Bukman and Shore
\cite{bs} as follows. They assumed that when $\gamma $ deviates from zero,
the contour becomes discontinuous at some point $x_{c}$. It is possible then
to solve Eq.(\ref{int_eq}) perturbatively considering the length of the gap
of the contour as a small parameter. It turns out that the solution exists
only if the break point $x_{c}$ satisfies the equation
\begin{equation}
R^{\left( 0\right) }(x_{c})=0,  \label{x_c}
\end{equation}%
which fixes the location of $x_{c}$ as well as the location of all the
contour $\Gamma $ in the limit $\gamma \rightarrow 0$. This method however
allows a calculation of only the first derivative of $\ \Lambda _{0}\left(
\gamma \right) $ at $\gamma =0$ in the thermodynamic limit, i.e. the leading
asymptotics of the average flow of particles. To study the behaviour of $%
\Lambda _{0}(\gamma )$ for nonzero values of $\gamma $, a calculation of the
finite size corrections to the above expression of $R^{\left( 0\right) }(x)$
is necessary. To this end, we use the method developed in \cite{kim,dl}. In
fact, once we have the expression for $R^{\left( 0\right) }(x)$, Eq.(\ref%
{R_0(x)}), it can be directly substituted to the formula for the finite size
asymptotic expansion of $R(x)$ obtained in \cite{pph2} \ for the ASAP. As a
result we obtain the following parametric dependence of \ $R\left( x\right) $
on $\gamma $, both being represented as the functions of the same parameter $%
\varkappa $,
\begin{eqnarray}
R_{s} &=&R_{s}^{\left( 0\right) }+\frac{1}{N^{3/2}}\frac{1}{2\pi i}\frac{%
q^{|s|}}{1-q^{|s|}}  \notag \\
&&\times \sum_{n=0}^{\infty }\left( \frac{i}{2N}\right) ^{n}\frac{\Gamma (n+%
\frac{3}{2})}{\pi ^{n+\frac{3}{2}}}\frac{c_{2n+1,s}}{\sqrt{2i}}\mathrm{Li}%
_{n+\frac{3}{2}}(\varkappa ) \\
\gamma  &=&-\frac{1}{N^{3/2}}\sum_{n=0}^{\infty }\left( \frac{i}{2N}\right)
^{n}\frac{\Gamma (n+\frac{3}{2})}{\pi ^{n+\frac{3}{2}}}\frac{\overline{c}%
_{2n+1}}{\sqrt{2i}}\mathrm{Li}_{n+\frac{3}{2}}(\varkappa )
\end{eqnarray}%
Here $\mathrm{Li}_{\alpha }(x)$ is the polylogarithm function,%
\begin{equation*}
\mathrm{Li}_{\alpha }(x)=\sum_{s=1}^{\infty }\frac{x^{s}}{s^{\alpha }},
\end{equation*}%
$R_{s}$ and $R_{s}^{\left( 0\right) }$ are the Laurent coefficients of $R(x)$
and $R^{\left( 0\right) }(x)$ respectively defined as follows
\begin{equation}
R(x)=\sum_{s=-\infty }^{\infty }R_{s}/x^{s+1},
\end{equation}%
and $c_{n,s}$ and $\overline{c}_{n}$ are the coefficients of $x^{n}$ in $%
\left( \sum_{k=0}^{\infty }a_{k}x^{k}\right) ^{s}$ and $\log \left(
\sum_{k=0}^{\infty }a_{k}x^{k}\right) $ respectively, where $a_{n}$ are the
coefficients of the inverse expansion $Z^{-1}(x)$ of the function $%
Z(x)=-\int_{x_{0}}^{x}R(x)dx$ near the point $Z(x_{c})$. As the derivative
of $Z(x)$ vanishes at $x=x_{c}$, Eq.(\ref{x_c}), the inverse expansion is
in powers of $\sqrt{x-Z(x_{c})}$,
\begin{equation*}
Z^{-1}(x)=\sum\limits_{n=0}^{\infty }a_{n}\left( x-Z(x_{c})\right) ^{n/2}.
\end{equation*}%
Being substituted to the Abel-Plana formula \cite{kim}, which is used to
evaluate the difference between the integral in Eq.(\ref{int_eq}) and the
sum in the original Bethe equations, it becomes a source of $1/\sqrt{N}$
corrections to $R^{\left( 0\right) }(x)$. The location of $x_{c}$ is to be
self-consistently defined from the equation $R\left( x_{c}\right) =0$. For
the first three orders of $1/\sqrt{N}$ expansion the\ coefficients $a_{n}$
can be obtained from the inverse expansion of zero order function $Z^{\left(
0\right) }(x)=-\int_{x_{0}}^{x}R^{\left( 0\right) }(x)dx$, while $R^{\left(
0\right) }(x)$ has been obtained above. The details of calculations can be
found in \cite{pph2}. The sum over the roots of a function $\mathrm{f}(x)$
analytic inside the contour can be evaluated in terms of the Laurent
coefficients $R_{s}$ of $R\left( x\right) $
\begin{equation}
\sum_{i=1}^{M}\mathrm{f}(x_{i})=2\pi iN\sum_{s=1}^{\infty }q^{-s}\mathrm{f}%
_{s}R_{s}.
\end{equation}%
and the of Taylor coefficients of $\mathrm{f}(x)$
\begin{equation}
\mathrm{f}(x)=\sum_{s=1}^{\infty }\mathrm{f}_{s}x^{s}.
\end{equation}%
This allows evaluation of $\ln \Lambda _{0}\left( \gamma \right) $. Finally
the point $x_{c}$ enters into all the results through the coefficients $%
c_{2n+1,s}$ and $\overline{c}_{2n+1}$. The final expression for $\Lambda
_{0}\left( \gamma \right) $ obtained in the scaling limit $\gamma
N^{3/2}=const$ \ has the form (\ref{ln Lamda_0}-\ref{x(C)}), which was
obtained earlier \ for the ASEP and the ASAP and claimed to be universal for
the KPZ universality class \cite{kpz,da}. The constants $\phi ,k_{1},k_{2}$
specific for the model under consideration,
\begin{eqnarray}
\phi  &=&\frac{\lambda x_{c}}{1+\lambda x_{c}},  \label{phi} \\
k_{1} &=&\sqrt{\frac{x_{c}}{8\pi }}\frac{2\lambda ^{2}g_{q,\lambda }^{\prime
}(x_{c})+\left( 1+\lambda x_{c}\right) \lambda g_{q,\lambda }^{\prime \prime
}(x_{c})}{\left( 1+\lambda x_{c}\right) ^{3}(g_{q,\lambda }^{\prime
}(x_{c}))^{5/2}},  \label{k1} \\
k_{2} &=&\sqrt{2\pi x_{c}g_{q,\lambda }^{\prime }\left( x_{c}\right) },
\label{k2}
\end{eqnarray}%
are expressed through the derivatives of the function $g_{q,\lambda }(x)$,
(Eq.(\ref{g(x)})). As it follows from the explicit form of Eq.(\ref{x_c}),
the unphysical parameter $x_{c}$ is related to the density $\rho $ by the
equation,
\begin{equation}
\rho =g_{q,\lambda }\left( x_{c}\right) .  \label{rho vs xc}
\end{equation}%
In Sec. VI we show that relations (\ref{phi},\ref{rho vs xc}) can be
obtained from the partition function formalism for the stationary state
without going into the Bethe ansatz solution. The parameter $x_{c}$ then
turns out to be related to the fugacity of a particle in the infinite
system. Thus, this result follows from the properties of the stationary
state only, rather than from the solution of the dynamical problem. In
general the partition function formalism allows the calculation of any
stationary spatial correlations. The Bethe ansatz, however, allows one to
probe into really dynamical characteristics, i.e. those, which are related
to unequal time correlation functions, such as higher cumulants of the
distance travelled by particles. They can be expressed through the constants
\ $k_{1},k_{2}$ from Eqs.(\ref{k1},\ref{k2}) by taking derivatives of the
formula (\ref{Lambda, q=0}), which yields the following behavior, specific
for the KPZ universality class,%
\begin{equation*}
\left\langle Y_{t}^{n}\right\rangle _{c}\sim N^{3\left( n-1\right)
/2}k_{1}k_{2}^{n}.
\end{equation*}

Studying the dependence of the results on the physical parameters $p$ and $q$
one should solve the equation (\ref{rho vs xc}) to find the behavior of \ $%
x_{c}$. Even without the explicit solution we can say that for generic
values of $p$ and $q$ \ the point $x_{c}$ is located at the positive part of
the real axis in the segment $[0,1)$, where the function $g_{q,\lambda
}\left( x\right) $ increases monotonously from $0$ to infinity. Apparently
the constants $\phi ,k_{1},k_{2}$ do not have any singularities for $x_{c}$
varying in this region and, thus, no abrupt change of the behavior is
expected. Particularly, the results obtained reproduce the corresponding
results for several models studied before. For example, in the continuous
time limit, $p\rightarrow 0$, keeping only the first order in $p$ we obtain
the corresponding constants for $q-$boson asymmetric diffusion model \cite%
{sw},\cite{pov}, which itself contains the drop push model \cite{srb} and
phase model \cite{bik} as particular cases, and can be related also to the
totally ASEP. The only interesting exception is the limiting case $%
p\rightarrow 1$, which turns out to exhibit a kind of phase transition
phenomena.

\section{ Asymmetric avalanche process and phase separation in the
deterministic limit.\label{ASAP}}

Let us first consider qualitatively what happens in the limit
\begin{equation}
p=1-\delta \tau ,\delta \tau \rightarrow 0,
\end{equation}%
in the domain
\begin{equation}
-1<q<0,\,\,\,\rho <1.
\end{equation}%
In this limit a particle from a single particle (SP) site, i.e. from a site
occupied by one particle only, almost definitely takes a step forward at
every time step. At the same time, one particle from a many particle (MP)
site, i.e. from a site occupied by more than one particle, takes a step with
the probability in general less than one. When the jumps of particles from\
SP sites are purely deterministic, i.e. the equality $p=1$ holds exactly,
the ZRP dynamics becomes "frozen" as soon as the system arrives at any
configuration consisting of SP sites only. By "frozen" we mean that at every
time step particles from all sites synchronously jump one step forward.
Thus, the structure of the configuration remains unchanged up to a uniform
translation. Therefore such configurations play the role of absorbing states
when $p=1$. When $p$ is less than $1$ by a small value $\delta \tau $, the
system can go from one absorbing state to another with the probability of
order of $\delta \tau $. This happens if at least one particle decides not
to jump.

It is easy to see that the limit under consideration is directly related to
the ASAP. Let us look at the system from the comoving reference frame, which
moves a step forward at every time step. In this frame particles from the SP
sites either stay with the probability $\left( 1-\delta \tau \right) $ or
take a step backward with the probability $\delta \tau $. At the same time
the MP sites play a role of \ columns (or avalanches) of particles moving
backward. Individual history of every such a column develops according to
the following rule. If $n>1$ particles occupy a site, either all $n$
particles jump to the next site backward with probability $\ \mu _{n}=\left(
1-p\left( n\right) \right) $ or $\left( n-1\right) $ particles jump and one
stays with probability $\left( 1-\mu _{n}\right) $. One can see that the
definition of probabilities $\mu _{n}$ coincides with those for the ASAP,
Eq.(\ref{mu_n}), in the leading order in $\delta \tau $. If we associate one
step of the discrete time $t$ with the continuous time interval $\delta \tau
$, neglect the terms smaller than $\delta \tau $ \ in the master equation
and take the limit $\delta \tau \rightarrow 0$, we come to the continuous
time master equation for the ASAP \cite{piph} . Note, that in the definition
of the ASAP at most one avalanche is allowed to exist at a time. In
principle, the discrete time dynamics allows for more than one \
simultaneous MP sites. However, the probability of formation of two or more
of them for a finite number of steps is of order of $\left( \delta \tau
\right) ^{2}$ and, as such, corresponding processes vanish when the limit $%
\delta \tau \rightarrow 0$ is taken.

As shown in \cite{piph} the ASAP exhibits the transition from the
intermittent to continuous flow regime. In this transition the average
avalanche size (the number of particle jumps in an avalanche) diverges in
the thermodynamic limit when the density of particles approaches the
critical point from below. To explain what this behaviour means in terms of
the discrete time model under consideration, we note that the average
avalanche size is roughly equal to the average number of particles occupying
an MP site times the average number of steps, which an MP site persists for.
Let us consider a configuration with one MP site on the lattice. In average,
the flow of particles into the MP site is equal to the density of mobile
(solitary) particles. The flow out of the MP site with $n$ particles is $%
p(n) $, which saturates exponentially rapidly to the limiting value $%
p(\infty )=1/(1-q)$ when $n$ grows. Therefore, if $\rho <\rho _{c}\equiv
1/(1-q)$, the outflow exceeds the inflow and the MP site tends to decay for
a few steps. As a result the average occupation number and the life time of
an MP site below the critical density are finite. On the contrary, if $\rho
>\rho _{c}$, the outflow is less than inflow and the dynamics favours the
growth of the MP sites. On the finite lattice the MP site grows at the
expense of the density of solitary particles until it absorbs enough
particles to equalize inflow and outflow. This happens when the MP site
contains $N(\rho -\rho _{c})$ particles. Any deviation from this number
destroys the balance between the inflow and outflow and the system tends to
restore this value again. As a consequence the life time of such an MP site
grows exponentially with the system size \cite{pph3}.

Of course the existence of only one MP site on the lattice is specific for
the continuous time ASAP, where the probability of appearance of two MP
sites simultaneously is neglible. In the discrete time case a possibility to
have many MP sites at the lattice should be considered as well. However, if
the characteristic occupation number of MP sites is such a large that the
outflow from MP sites can be roughly replaced by $p(\infty )$, the above
claim based on the balance between inflow and outflow remains valid: the
total number of particles in all MP sites is equal to $N(\rho -\rho _{c})$.
As a result, the particle current growing linearly with the density below
the critical point, stops growing at $\rho _{c}$ due to appearance of MP
sites, the occupation numbers of which grow instead. Of course these
arguments describe the limiting situation $p\rightarrow 1$. When $\delta
\tau $ is finite, the transition is smoothened. As shown in Fig.(\ref%
{phiplot}), the flow-density plot, obtained by substituting the result of
numerical solution of Eq.(\ref{rho vs xc}) to Eq.(\ref{phi}), approaches the
form of broken line consisting of two straight segments: $\phi =\rho $ for $%
\rho <$ $\rho _{c}$ and $\phi =\rho _{c}$ for $\rho \geq $ $\rho _{c}$.
\begin{figure}[tbp]
\unitlength=1mm \makebox(110,45)[cc] {\psfig{file=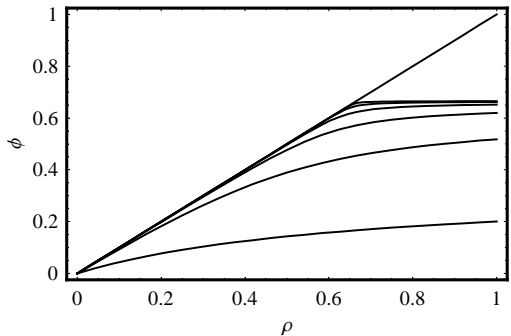,width=70mm}}
\caption{The flow-density plots for $q=-1/2$ and $\protect\delta \protect%
\tau =0.5,0.5\times 10^{-1},0.5\times 10^{-2},0.5\times 10^{-3},0.5\times
10^{-4},0.5\times 10^{-5},0,$ going in bottom-up direction respectively.}
\label{phiplot}
\end{figure}

Of course the above qualitative arguments being far from rigorous can only
serve as an illustration. A specific distribution of the occupation numbers
of MP sites can be obtained with the help of the canonical partition
function formalism, which is described in details in the next section. In
present section we analyse the behaviour of generating function $\Lambda
_{0}(\gamma )$ of the cumulants of the integrated particle current $Y_{t}$
that can be extracted form the solution of the Bethe ansatz equations
discussed above.

To this end, we note that the universal scaling form (\ref{ln Lamda_0}-\ref%
{x(C)}) of $\Lambda _{0}(\gamma )$ holds in all the phase space, provided
that the parameters $\phi ,k_{1},k_{2}$ are finite. What depends on the
parameters of the problem is the typical scales encoded in $\phi
,k_{1},k_{2} $, which characterize the average and the fluctuations of the
integrated particle current $Y_{t}$.

Before going to the analysis let us look how the results are related to
those for the ASAP. To this end, we recall that in order to transform the
ZRP to the ASAP we consider comoving reference frame, which takes one step
forward at each time step. The distances, which particles travel in these
reference frames, are related to each other as follows
\begin{equation}
Y_{t}^{ZRP}=Mt-Y_{t}^{ASAP}.
\end{equation}%
Hence, the largest eigenvalues of the equations for the generating functions
$F_{t}\left( C\right) $ for the ASAP and ZRP are related as follows%
\begin{equation}
\ln \Lambda _{0}\left( \gamma \right) =\gamma M-\delta \tau \lambda
_{0}^{ASAP}+O(\delta \tau ^{2}).  \label{ln lambda (p=1)}
\end{equation}%
By $\lambda _{0}^{ASAP}$ we mean the largest eigenvalue of the continuous
time equation for the ASAP obtained in \cite{pph2}. The factor $\delta \tau $
which comes with $\lambda _{0}^{ASAP}$ reflects the time rescaling, in which
one discrete time step is identified with the infinitesimal interval $\delta
\tau $. The formula (\ref{ln lambda (p=1)}) suggests that the expansion of $%
\ln \Lambda _{0}\left( \gamma \right) $ in powers of $\delta \tau $ exists,
which turns out to be true only in the subcritical region $\rho <\rho _{c}$.
In what follows we solve the equation (\ref{rho vs xc}) in the limit $%
p\rightarrow 1$, i.e. $\delta \tau \rightarrow 0$. Once the value of $x_{c}$
is obtained from this solution, it is to be directly substituted to the
formulas (\ref{phi}-\ref{k2}) to obtain $\phi ,k_{1},k_{2}$.

When $\delta \tau =0$ the function $g_{q,\lambda }\left( x\right) $ can be
summed up to the following simple form%
\begin{equation}
g_{q,-1/q}\left( x\right) =\frac{x}{x-q},  \label{g_q,-1/q(x)}
\end{equation}%
such that Eq.(\ref{rho vs xc}) $\ $can be easily solved
\begin{equation}
x_{c}=\frac{q\rho }{\rho -1}.  \label{x_c_1}
\end{equation}%
Then Eqs.(\ref{phi}-\ref{k2}) yield: $\phi =$ $\rho $ and $k_{1}=0$, i.e. $%
\ln \Lambda _{0}=\gamma M$. This corresponds to synchronous jumps of all
particles without any stochasticity. For small nonzero $\delta \tau $ we
expect the correction of order of $\delta \tau $ to appear. This should be
true at least if the argument of $g_{q,\lambda }\left( x\right) $ is away
from any of its singularities, which turns out to be the case only in the
subcritical region $\rho <\rho _{c}$. In general in this case one can
represent $g_{q,\lambda }\left( x\right) $ as a Taylor series in powers of $%
\delta \tau $ with non-singular series coefficients. Then \ Eq.(\ref{rho vs
xc}) can be solved perturbatively order by order in $\delta \tau $. To this
end, we look for the solution in the form of the perturbative expansion in
powers of $\ \delta \tau $
\begin{equation}
x_{c}=x_{c}^{(0)}+\delta \tau x_{c}^{(1)}+\cdots
\end{equation}%
The first order solution of Eq.(\ref{rho vs xc}) yields
\begin{equation}
x_{c}^{(1)}=\frac{\left( 1-q\right) \rho }{\left( 1-\rho \right) }\left(
1-\left( \sum\limits_{s=1}^{\infty }\frac{q^{s}}{\left( 1+(q^{s}-1)\rho
\right) ^{2}}\right) \right) ,
\end{equation}%
which results in the following values of $\phi ,k_{1},k_{2}$:
\begin{eqnarray}
\phi  &\simeq &\rho +\delta \tau \left[ \frac{\left( 1-q\right) }{q}\left(
\sum\limits_{s=1}^{\infty }\frac{s}{1-q^{s}}\left( \frac{q\rho }{\rho -1}%
\right) ^{s}\right) \right] , \\
k_{1} &\simeq &\delta \tau \frac{\left( q-1\right) }{\sqrt{8\pi }q\left(
(1-\rho )\rho \right) ^{3/2}}\sum\limits_{s=1}^{\infty }\frac{%
s^{2}(s-1+2\rho )}{1-q^{s}}\left( \frac{q\rho }{\rho -1}\right) ^{s}, \\
k_{2} &\simeq &\sqrt{2\pi (1-\rho )\rho }.
\end{eqnarray}%
Thus, the first order calculation allows taking into account the
fluctuations of the particle current due to spontaneous MP site formation.
The values obtained indeed reproduce $\lambda _{0}^{ASAP}$ from \cite{pph2}.
Close to the critical point $x_{c}^{(1)}$ diverges,
\begin{equation*}
x_{c}^{(1)}\sim \left( \rho _{c}-\rho \right) ^{-2},
\end{equation*}%
resulting in the divergent contributions to $\phi ,k_{1},k_{2}$
\begin{equation*}
\phi ^{(1)}\sim \left( \rho _{c}-\rho \right) ^{-2},k_{1}^{(1)}\sim \left(
\rho _{c}-\rho \right) ^{-4},k_{2}^{(1)}\sim \left( \rho _{c}-\rho \right)
^{-3}.
\end{equation*}%
The reason of these divergencies have a clear physical meaning. For example,
the first order correction to the flow $\phi $ gives a fraction of
particles, which stay in MP sites. Roughly speaking it is a product, $%
t_{MP}n_{MP}p_{MP}$, of the life time of an MP site $t_{MP}$, its average
occupation number $n_{MP}$ and the probability of appearance $p_{MP}$, the
latter being of order of $\ \delta \tau $. As it follows from the above
arguments based on the balance between the flows into and out of an MP site,
the two former, $t_{MP}$ and $n_{MP}$, should diverge as the density of
particles approaches its critical value.

The singularities in the solution limit the applicability of the
perturbative scheme used. It to be applicable the solution $x_{c}$ should be
far enough from the first singularity of $g_{q,\lambda }\left( x\right) $,
at the positive half-axis, $x=1$. Specifically, the parameter $\left( \delta
\tau x_{c}^{(1)}/(\rho _{c}-\rho )\right) $ should be small to ensure the
existence of nonsingular expansion for $g_{q,\lambda }\left( x\right) $.
This yields
\begin{equation}
\delta \tau (\rho _{c}-\rho )^{-3}\ll 1,
\end{equation}%
resulting in%
\begin{equation}
1-x_{c}\sim \rho _{c}-\rho \gg \delta \tau ^{1/3}.  \label{1-x_c}
\end{equation}

To proceed further, we note that the function $g_{q,\lambda }\left( x\right)
$ monotonously grows along the positive half-axis from zero at the origin, $%
x=0$, to infinity at the pole $x=1$. As the function $g_{q,\lambda }\left(
x\right) $ runs over all positive real values between these two points,
including the value of $\rho $, the solution of Eq.(\ref{rho vs xc}), $x_{c}$%
, is always located at the segment $0\leq x\leq 1$. \ According to the above
perturbative solution almost all this segment is exhausted by the
subcritical domain, Eq.(\ref{1-x_c}), except the small vicinity of $x=1$,
i.e. $\left( 1-x_{c}\right) \lesssim \delta \tau ^{1/3}$. Therefore, this
vicinity is where $x_{c}$ should be looked for at the critical point and
above.

This remark can be used as a basis for another perturbative scheme. Let us
suppose that
\begin{equation}
x_{c}=1-\Delta ,  \label{x_c=1-Delta}
\end{equation}%
where $\Delta $ is a small parameter. The function $g_{q,\lambda }\left(
x\right) $ can be represented in the following form
\begin{equation}
g_{q,\lambda }\left( x\right) =\sum_{k=0}^{\infty }\frac{\lambda xq^{k}}{%
1+q^{k}\lambda x}+\sum_{k=0}^{\infty }\frac{q^{k}x}{1-q^{k}x}.
\label{g(x)_1}
\end{equation}%
This is obtained by expanding the expression under the sum in Eq.(\ref{g(x)}%
) in powers of $\ q^{n}$ and then changing the order of summations. In this
form $g_{q,\lambda }\left( x\right) $ has a form of the sum of terms, each
with one of the following simple poles%
\begin{equation*}
x_{k}^{\prime }=q^{-k},\qquad x_{k}^{\prime \prime }=-q^{-k}\lambda
^{-1},\qquad k=0,1,2,\ldots
\end{equation*}%
The pairs of the poles $x_{k}^{\prime }$ and $x_{k-1}^{\prime \prime }$
merge when $\delta \tau $ tends to zero, so that corresponding terms having
equal absolute values and opposite signs cancel each other. As a result, for
finite $\Delta $ the only term of $g_{q,\lambda }\left( x_{c}\right) $,
which survives in the limit $\delta \tau \rightarrow 0,$ is the one with the
pole $x_{0}^{\prime \prime }$ (see Eq.(\ref{g_q,-1/q(x)})). On the other
hand the term under the first sum in Eq.(\ref{g(x)_1})\ with the pole $%
x_{0}^{\prime }=1$ tends to infinity when $\Delta $ tends to zero. The
compromise is to consider the two limits simultaneously. Then the divergent
contribution from the pole $x_{0}^{\prime }$ can be cancelled up to a finite
constant by the term with the pole $x_{1}^{\prime \prime }=-\left( q\lambda
\right) ^{-1}$. The value of the resulting constant, which depends on the
relation between $\delta \tau $ and $\Delta $, should be chosen such as to
satisfy Eq.(\ref{rho vs xc}). Substitution of Eq.(\ref{x_c=1-Delta}) to Eq.(%
\ref{rho vs xc}) shows that the value of $g_{q,\lambda }\left( x_{c}\right) $
to be finite in the limit $\Delta $ $\rightarrow 0$, the following ratio
should be kept constant.
\begin{equation}
\delta \tau /\Delta ^{2}\rightarrow const
\end{equation}%
Using this fact we can look for the relation between $\delta \tau $ and $%
\Delta $ in form of the expansion in $\Delta $%
\begin{equation}
\delta \tau =\left( 1-1/q\right) ^{-1}\left( \alpha \Delta ^{2}+\beta \Delta
^{3}+O(\Delta ^{4})\right) .  \label{delta tau}
\end{equation}%
The coefficient $\left( 1-1/q\right) ^{-1}$ is for the further convenience.
Solving the equation (\ref{rho vs xc}) order by order in powers of $\Delta $
we fix $\alpha $ and $\beta $,
\begin{eqnarray}
\alpha &=&\rho -\rho _{c}  \label{a} \\
\beta &=&\alpha (1+\alpha )-q\left( 1-q\right) ^{-2}.  \label{b}
\end{eqnarray}%
Inverting the relation (\ref{delta tau}) we express $x_{c}$ in terms of $%
\delta \tau $. Finally we substitute $x_{c}$ into Eqs.(\ref{phi}-\ref{k2}).
The cases $\alpha =0$ and $\alpha >0$ should be considered separately.

-- \textit{In the case }$\alpha >0$\textit{, which corresponds to the
density above critical, }$\rho >\rho _{c}$\textit{,} we obtain in the
leading order
\begin{eqnarray}
\phi &\simeq &\rho _{c}-\delta \tau ^{1/2}\left[ \frac{\rho _{c}^{2}\left(
1-\rho _{c}\right) }{\rho -\rho _{c}}\right] ^{1/2}, \\
k_{1} &\simeq &\frac{3}{8\sqrt{\pi }}\rho _{c}\left( 1-\rho _{c}\right)
^{3/4}\frac{\delta \tau ^{1/4}}{\left( \rho -\rho _{c}\right) ^{7/4}}, \\
k_{2} &\simeq &2\sqrt{\pi }\left( 1-\rho _{c}\right) ^{1/4}\frac{\left( \rho
-\rho _{c}\right) ^{3/4}}{\delta \tau ^{1/4}}.
\end{eqnarray}%
As follows form these formulas the flow of particles $\phi $ in the
thermodynamic limit is equal to the critical density $\rho _{c}$ as
expected. It is possible also to obtain $1/N$ correction to the
thermodynamic value of the particle flow $\phi $, which was shown to be
universal being proportional to the nonlinear coefficient $\lambda $ of the
KPZ equation \cite{Krug1}.%
\begin{equation}
\phi _{N}-\phi \simeq \frac{1}{N}\frac{3}{4}\frac{\rho _{c}\left( 1-\rho
_{c}\right) }{\left( \rho -\rho _{c}\right) }
\end{equation}%
The value of the correction diverges in the critical point. Surprisingly its
value does not depend on $\delta \tau $. Higher cumulants, responsible for
the fluctuations around the average flow grow when $\delta \tau $ tends to
zero,
\begin{equation}
\lim_{t\rightarrow \infty }\frac{\left\langle Y_{t}^{n}\right\rangle _{c}}{t}%
\sim N^{3\left( n-1\right) /2}\frac{\left( \rho -\rho _{c}\right) ^{\left(
3n-7\right) /4}}{\delta \tau ^{\left( n-1\right) /4}}.
\end{equation}

-- \textit{In the critical point }$\alpha =0$\textit{, i.e. }$\rho =\rho _{c}
$\textit{,} we obtain,
\begin{eqnarray}
\phi _{c} &\simeq &\rho _{c}-\delta \tau ^{1/3}\left[ \rho _{c}^{2}\left(
1-\rho _{c}\right) \right] ^{1/3}, \\
k_{1} &\simeq &\left( \frac{\rho _{c}}{\delta \tau }\right) ^{1/3}\left[
54\pi \left( 1-\rho _{c}\right) \rho _{c}\right] ^{-1/2}, \\
k_{2} &\simeq &\sqrt{6\pi \left( 1-\rho _{c}\right) \rho _{c}}.
\end{eqnarray}%
The $1/N$ correction to the average flow looks as follows
\begin{equation}
\phi _{N}-\phi \simeq \frac{1}{N}\frac{1}{3}\left( \frac{\rho _{c}}{\delta
\tau }\right) ^{1/3},
\end{equation}%
growing with the decrease of $\delta \tau .$ The other cumulants have
similar $\delta \tau $ behavior,%
\begin{equation}
\lim_{t\rightarrow \infty }\frac{\left\langle Y_{t}^{n}\right\rangle _{c}}{t}%
\sim N^{3\left( n-1\right) /2}\delta \tau ^{-1/3}.
\end{equation}%
As follows from the results obtained, though the average flow saturates to a
constant value at the critical point, its fluctuations grow when $\delta
\tau $ goes to zero.

The generating function of cumulants $\Lambda _{0}(\gamma )$ can be related
to the large deviation function for the particle current \cite{dl}. As a
result the probability distribution \ of $Y_{t}/t$ reads as follows

\begin{equation}
P\left( \frac{Y_{t}}{t}=y\right) =\exp \left[ -\frac{t}{t_{c}}H\left( \frac{%
y-N\phi }{\mathcal{G}}\right) \right] ,  \label{ldf}
\end{equation}%
where the universal function $H(x)$ is given by the following parametric
expression
\begin{eqnarray}
H(x) &=&\beta G^{\prime }\left( \beta \right) -G\left( \beta \right) \\
x &=&G^{\prime }(\beta ).
\end{eqnarray}%
with two model dependent constants%
\begin{eqnarray*}
t_{c} &=&N^{3/2}k_{1}^{-1} \\
\mathcal{G} &=&k_{1}k_{2}.
\end{eqnarray*}%
The function $H(x)$ behaves as $\left( x-1\right) ^{2}$ for $x\ll 1$ and as $%
x^{5/2}$ and $x^{3/2}$ as $x$ goes to plus and minus infinity respectively.
One can see from Eq.(\ref{ldf}) that the parameter $\mathcal{G}$ plays the
role of the characteristic scale at which the current fluctuations become
non-Gaussian. This takes place when the argument of $H(x)$ in Eq.(\ref{ldf})
becomes of order of $1$. Note that $\mathcal{G}$ coincides with $1/N$
correction to the average flow. This is a reflection of the fact that the
correlations between particle jumps, which cause deviations from the
Gaussian behaviour, owe to the finiteness of the system and to the periodic
boundary conditions. The parameter $t_{c}$ is the characteristic time in
which such fluctuations become unlikely. The characteristic time $t_{c}$
being of order of $N^{3/2}$ signifies the KPZ behaviour characterized by the
dynamical exponent $z=3/2$. In fact, for the totally \cite{gs} and the
partially \cite{kim} ASEP the inverse time $t_{c}^{-1}$ coincides with the
next to the largest eigenvalue of the master equation up to a constant of
order of 1. We expect this to hold also in our case. Below the critical
point $\rho <\rho _{c}$ we have $t_{c}\sim N^{3/2}\delta \tau ^{-1}$ and $%
\mathcal{G}\sim \delta \tau $. Above the critical point, $\rho >\rho _{c}$,
the characteristic time scales as $t_{c}\sim N^{3/2}\delta \tau ^{-1/4}$
growing with the decay of $\delta \tau $ and the fluctuation scale is finite
$\mathcal{G}\sim \left( \rho -\rho _{c}\right) ^{-1}$ growing as the density
approaches the critical point. Exactly at the critical point we obtain $%
\mathcal{G}\sim \delta \tau ^{-1/3}$ and $t_{c}\sim \delta \tau
^{1/3}N^{3/2} $. One can see that below the critical point even very small
fluctuations are non-Gaussian. Above the critical point the range of
Gaussian fluctuations is finite. At the critical point the fluctuations
remain Gaussian in a very wide range. On the other hand the \ time of decay
of the fluctuations at the critical point is much smaller than below and
above.

\section{Canonical partition function formalism for the stationary state.
\label{Partition function}}

In this section we show that, the partition function formalism \cite{evans},%
\cite{Burda} allows a calculation of some physical quantities yielding the
same results as obtained above. Particularly the results obtained in the
saddle point approximation are equivalent to the results obtained from the
above solution of the Bethe equations in the thermodynamic limit. We also
analyze the change of the occupation number distribution under the phase
transition in the limit $\delta \tau \rightarrow 0$.

\subsection{Canonical partition function and stationary correlations.}

The partition function of the ZRP is defined as the normalization constant
of the stationary distribution (\ref{P_st})%
\begin{equation}
Z\left( N,M\right) =\sum_{\left\{ n_{i}\right\} }\prod_{i=1}^{N}f\left(
n_{i}\right) \delta \left( n_{1}+\cdots +n_{N}\right) ,
\end{equation}%
where $f(n)$ is the one-site weight defined in Eq.(\ref{f(n)}). The sum can
be represented in the form of the contour integral,%
\begin{equation}
Z\left( N,M\right) =\oint \frac{\left( F(z)\right) ^{N}}{z^{M+1}}\frac{dz}{%
2\pi i},  \label{Z(N,M)}
\end{equation}%
where $F(z)$ is the generating function of the one-site weights,
\begin{equation}
F\left( z\right) =\sum_{n=0}^{\infty }z^{n}f\left( n\right) .
\end{equation}%
Once the partition function is known, it can be used to obtain stationary
correlation functions. For example the probability $P\left( n\right) $ for a
site to hold $n$ particles is as follows,
\begin{equation}
P\left( n\right) =f\left( n\right) \frac{Z\left( N-1,M-n\right) }{Z\left(
N,M\right) }\text{.}  \label{P(n)=Z/Z}
\end{equation}%
Another tool characterizing the occupation number distribution is the
generating function of its moments $\left\langle e^{\gamma n}\right\rangle $%
, which can be also represented in the form of contour integral%
\begin{eqnarray}
\left\langle e^{\gamma n}\right\rangle &=&\sum\limits_{k=0}^{\infty }\frac{%
\gamma ^{k}\left\langle n^{k}\right\rangle }{k!}  \label{<e^gamma n>} \\
&=&\frac{1}{Z\left( N,M\right) }\oint \frac{\left( F(z)\right) ^{N-1}F\left(
ze^{\gamma }\right) }{z^{M+1}}\frac{dz}{2\pi i}.  \notag
\end{eqnarray}%
Another quantity of interest is the probability $\mathcal{P(}n\mathcal{)}\ $%
for $n$ particles to hop simultaneously. The integral representation of the
corresponding generating function is
\begin{equation}
\Psi \left( x\right) \equiv \sum_{n=0}^{M}x^{n}\mathcal{P(}n)\mathcal{=}%
\frac{1}{Z\left( N,M\right) }\oint \frac{\left[ \Phi \left( x,z\right) %
\right] ^{N}}{z^{M+1}}\frac{dz}{2\pi i},  \label{Psi(x)}
\end{equation}%
where%
\begin{equation*}
\Phi \left( x,z\right) =\sum_{n=0}^{\infty }f\left( n\right) z^{n}\left(
xp\left( n\right) +\left( 1-p(n)\right) \right) .
\end{equation*}%
This yields for example the following expression for the average flow of
particles $\phi _{N}$,
\begin{equation}
\phi _{N}=\frac{1}{N}\Psi ^{\prime }\left( 1\right) =\frac{1}{Z\left(
N,M\right) }\oint \frac{\left[ F(z)\right] ^{N}}{z^{M+1}}\frac{z}{1+z}\frac{%
dz}{2\pi i}.  \label{phi int}
\end{equation}%
The subscript $N\,\ $specifies that the expression is valid for an arbitrary
finite lattice.

\subsection{The saddle point approximation.}

The integral in Eq.(\ref{Z(N,M)}) can be evaluated in the saddle point
approximation. The location of the saddle point $z^{\ast }$ is defined by
the following equation.
\begin{equation}
\rho =z^{\ast }\frac{F^{\prime }(z^{\ast })}{F(z^{\ast })}.
\label{saddle point (general)}
\end{equation}%
This equation can be treated as an equation of state, which relates the
density of particles $\rho $ and the fugacity of a particle $z^{\ast }$. It
is convenient to introduce the Helmholtz free energy%
\begin{eqnarray}
A(N,M) &=&-\ln Z(N,M) \\
&\simeq &-N\ln F(z^{\ast })+M\ln z^{\ast }.  \notag
\end{eqnarray}%
Then one can define the chemical potential
\begin{equation}
\mu =\left. \frac{\partial A(N,M)}{\partial M}\right\vert _{N}=\ln z^{\ast }
\end{equation}%
and the analogue of pressure
\begin{equation}
\mathcal{\pi }=-\left. \frac{\partial A(N,M)}{\partial N}\right\vert
_{M}=\ln F(z^{\ast }).
\end{equation}%
The stationary correlation functions mentioned above can also be expressed
in terms of the values defined.

-\textit{The occupation number probability distribution }$P(n)$ \textit{for}
$n\ll N$%
\begin{equation}
P(n)=f(n)e^{n\mu -\mathcal{\pi }}.  \label{P(n)}
\end{equation}%
We should note that when $n$ is of order of $N$ this equation loses the
validity as one should take into account the shift of the saddle point in
the integral for $Z(N-1,M-n)$. In other words, when $n$ in given site is
large it changes the density of particles in the other sites, which in its
turn leads to the change of the fugacity of particles.

-\textit{The cumulants }$\left\langle n^{k}\right\rangle _{c}$ \textit{\ of
the occupation number }$n$ are given by the derivatives of $\log
\left\langle e^{\gamma n}\right\rangle $, Eq.(\ref{<e^gamma n>}),\textit{\ }%
which leads to a standard relation between fluctuations and pressure
\begin{equation}
\left\langle n^{k}\right\rangle _{c}=\frac{\partial ^{k}\mathcal{\pi }}{%
\partial \mu ^{k}}.  \label{log <e^ngamma>}
\end{equation}

\textit{-The average flow of particles} $\phi $ is given by%
\begin{equation}
\phi =\frac{e^{\mu }}{1+e^{\mu }}.  \label{phi_1}
\end{equation}%
The$\ 1/N$ correction to the average flow can be given in terms of two first
cumulants of the occupation number $\left\langle n^{2}\right\rangle _{c}$, $%
\left\langle n^{3}\right\rangle _{c}$,%
\begin{equation}
\phi _{N}-\phi \simeq \frac{1}{2N}\left( \frac{\partial \phi }{\partial \mu }%
\frac{\left\langle n^{3}\right\rangle _{c}}{\left\langle n^{2}\right\rangle
_{c}^{2}}-\frac{\partial ^{2}\phi }{\partial \mu ^{2}}\frac{1}{\left\langle
n^{2}\right\rangle _{c}}\right) ,  \label{phi_n-phi}
\end{equation}%
which in its turn can be reduced to a very simple form
\begin{equation}
\phi _{N}-\phi =-\frac{1}{2N}\frac{d}{d\mu }\left( \frac{d\phi }{d\rho }%
\right) ,
\end{equation}%
where one should take into account the equation of state (\ref{saddle point
(general)}). The $1/N$ correction to the average flow has been claimed to be
universal for the KPZ class, depending only on the parameters of the
corresponding\ KPZ equation. Below we use it to reexpress the above
parameters $k_{1}$,$k_{2}$ obtained from the solution of the dynamical
problem in terms of the parameters of the stationary state only.

We should note that the criterion of validity of the saddle point
approximation, i.e. smallness of the second term of the asymptotical
expansion comparing to the first, yields the following range of parameters:%
\begin{equation}
\frac{\left\langle n^{3}\right\rangle _{c}^{2}}{\left\langle
n^{2}\right\rangle _{c}^{3}}\ll \sqrt{N}.  \label{criterion}
\end{equation}

Let us turn to the examples.

-- \textit{For the case }$q=0$, the series $F(z)$ expressed in terms of the
parameter $\lambda =p/(1-p)$ can be summed up to the form
\begin{equation}
F(z)=\frac{1}{1+\lambda }+\frac{z}{\lambda -z}.
\end{equation}%
In this case the integral in (\ref{Z(N,M)}) can be evaluated explicitly
resulting in%
\begin{equation}
Z\left( N,M\right) =\lambda ^{-M}\left( 1+\lambda \right) ^{-N}\frac{\Gamma
\left( L\right) }{\Gamma \left( N\right) \Gamma \left( M+1\right) }\left.
_{2}F_{1}\right. \left(
\begin{array}{c}
-M,-N \\
1-L%
\end{array}%
;-\lambda \right) .
\end{equation}%
Using then the formula (\ref{phi int}) we arrive at the expression of the
particle flow $\phi $ given in (\ref{vexact}).

-- \textit{In the case of arbitrary }$q,$ $|q|<1,$ due to specific form of
the one site weights $f(n)$, the function $F(z)$ has the structure of well
known $q-$series,
\begin{equation}
F(z)=\left( 1-p\right) \sum_{n=0}^{\infty }\left( \frac{z}{\lambda }\right)
^{n}\frac{\left( -\lambda ;q\right) _{n}}{\left( q;q\right) _{n}},
\label{F(z)}
\end{equation}%
where
\begin{equation}
(a;q)_{n}=\prod_{k=0}^{n-1}\left( 1-aq^{k}\right)
\end{equation}%
is a notation for shifted $q-$factorial. The series (\ref{F(z)}) can be
summed up to the infinite product using the $q-$binomial theorem \cite{aar},
which is the $q-$analog of the Taylor series of a power law function,%
\begin{equation}
F(z)=\left( 1-p\right) \frac{\left( -z;q\right) _{\infty }}{\left( z/\lambda
;q\right) _{\infty }}.  \label{q-binomial}
\end{equation}%
This formula (\ref{q-binomial}) turns out to be extremely useful as it
allows one to write the equation for the saddle point (\ref{saddle point
(general)}) in the following simple form.
\begin{equation}
\rho =\sum_{k=0}^{\infty }\frac{z^{\ast }q^{k}}{1+q^{k}z^{\ast }}%
+\sum_{k=0}^{\infty }\frac{z^{\ast }q^{k}/\lambda }{1-q^{k}z^{\ast }/\lambda
}  \label{saddle point}
\end{equation}%
Note that if we define%
\begin{equation}
x_{c}=z^{\ast }/\lambda ,  \label{x_c=z/labda}
\end{equation}%
then the r.h.s of \ Eq.(\ref{saddle point}) \ coincides with $g_{q,\lambda
}\left( x_{c}\right) $ in the form (\ref{g(x)_1}). Thus, the equation for
the saddle point is nothing but the conical point equation, Eq.(\ref{rho vs
xc}), appeared in the thermodynamic solution of the Bethe ansatz equations,
while the above parameter $x_{c}$ is proportional to the fugacity of a
particle $z^{\ast }$.

Using the equality
\begin{equation*}
g_{q,\lambda }\left( z/\lambda \right) =z\frac{F^{\prime }(z)}{F(z)}
\end{equation*}%
and Eq.(\ref{log <e^ngamma>}) one can obtain the explicit form \ of the
first two cumulants of the occupation \ number%
\begin{eqnarray}
\left\langle n^{2}\right\rangle _{c} &=&\frac{z^{\ast }}{\lambda }%
g_{q,\lambda }^{\prime }\left( z^{\ast }/\lambda \right) ,  \label{<n_c^2>}
\\
\left\langle n^{3}\right\rangle _{c} &=&\frac{z^{\ast }}{\lambda }\left(
\frac{g_{q,\lambda }^{\prime }\left( z^{\ast }/\lambda \right) }{\lambda }+%
\frac{z^{\ast }g_{q,\lambda }^{\prime \prime }\left( z^{\ast }/\lambda
\right) }{\lambda ^{2}}\right) .  \label{<n_c^3>}
\end{eqnarray}%
Substituting them into Eqs.(\ref{phi_1},\ref{phi_n-phi}) one can make sure
that the expression for the particle flow coincides with that obtained from
the Bethe ansatz solution.

It is remarkable to note that the variance of the occupation number, $%
\left\langle n^{2}\right\rangle _{c}$, can be directly related to the
parameter $k_{2}$,
\begin{equation}
k_{2}=\sqrt{2\pi \left\langle n^{2}\right\rangle _{c}}=\sqrt{2\pi \frac{%
\partial \rho }{\partial \mu }}.
\end{equation}%
At the same time the $1/N$ correction to the average flow, Eq.(\ref%
{phi_n-phi}), is given by the product $k_{1}k_{2}$ such that \ \ \
\begin{eqnarray}
k_{1}k_{2} &=&\frac{1}{2N}\left( \frac{\partial \phi }{\partial \mu }\frac{%
\left\langle n^{3}\right\rangle _{c}}{\left\langle n^{2}\right\rangle
_{c}^{2}}-\frac{\partial ^{2}\phi }{\partial \mu ^{2}}\frac{1}{\left\langle
n^{2}\right\rangle _{c}}\right)  \notag \\
&=&-\frac{1}{2}\frac{d}{d\mu }\left( \frac{d\phi }{d\rho }\right) .
\end{eqnarray}%
Thus, we have expressed the parameters characterizing the fluctuations of
the particle current in terms of the parameters of the stationary state,
which characterize the fluctuations of occupation numbers. As the obtaining
of the latter does not requires an integrability, we expect this relation to
hold for the general discrete time ZRP belonging to the KPZ class.

Another interesting relation can be found. Let us consider the free energy
per site, $a(\rho ,z)$ as a function of the density $\rho $ and the fugacity
$z$ that formally can take on arbitrary complex values%
\begin{equation*}
a(\rho ,z)=\lim_{N\rightarrow \infty }\frac{A(M,N)}{N}=-\ln F(z)+\rho \ln z.
\end{equation*}%
\ Then the following relation holds%
\begin{equation*}
\frac{\partial a(\rho ,z)}{\partial z}=\frac{1}{2\pi i}R_{0}\left( z\right) ,
\end{equation*}%
where $R_{0}\left( z\right) $ is the density of the Bethe roots in the
thermodynamic limit, Eq.(\ref{R_0(x)}). Thus the difference $\left\vert
a(\rho ,z_{1})-a(\rho ,z_{2})\right\vert $ with the values of $z_{1},z_{2}$
taken at the contour of the Bethe roots gives the fraction of the roots in
the segment between $z_{1}$and $z_{2}$. We should also note that the
equation of state, which in terms of $a(\rho ,z)$ looks as follows
\begin{equation*}
\partial a(\rho ,z)/\partial z=0,
\end{equation*}%
is equivalent to the conical point equation Eq.(\ref{x_c}).

\subsection{The occupation number distribution in the deterministic limit.}

As we noted above, the saddle point equation, Eq.(\ref{saddle point}), for
the integral that represents the partition function $Z(M,N)$, Eq.(\ref%
{Z(N,M)}), coincides with Eq.(\ref{rho vs xc}) appeared in the thermodynamic
solution of the Bethe equations. Therefore, we can directly use the results
of the solution of this equation obtained in the section \ref{ASAP} to
obtain the occupation number distribution in the limit $\delta \tau
\rightarrow 0$.

\textit{-- In the domain }$\rho <\rho _{c}$\textit{\ and }$n\ll N$\textit{\
we have}%
\begin{eqnarray}
P(0) &=&1-\rho -O(\delta \tau ),\text{ }P(1)=\rho -O(\delta \tau ) \\
P(n) &=&\delta \tau \frac{\left( 1-\rho \right) \left( 1-\rho _{c}\right)
^{-2}}{\left( 1-q^{n}\right) \left( 1-q^{n-1}\right) }\left[ \frac{\left(
1-\rho _{c}\right) }{\left( 1-\rho \right) }\frac{\rho }{\rho _{c}}\right]
^{n}+O(\delta \tau ^{2}),n\geq 2
\end{eqnarray}%
For $n\gg 1$ the latter formula yields the exponential decay.
\begin{equation*}
P(n)\sim \delta \tau \frac{\left( 1-\rho \right) }{\left( 1-\rho _{c}\right)
^{2}}\exp \left[ -n\frac{\rho _{c}-\rho }{\left( 1-\rho \right) \rho _{c}}%
\right]
\end{equation*}%
When the density approaches the critical point from below, $\left( \rho
_{c}-\rho \right) \ll 1$, the cut-off of this distribution diverges
proportionally to $\left( \rho _{c}-\rho \right) ^{-1}$.Note that strictly
below $\rho _{c}$ the cut-off is always finite, being independent of $\delta
\tau $, while $P(n)$ itself is of order of $\delta \tau $.

\textit{-- In the domain }$\rho >\rho _{c}$\textit{\ and }$n\ll N$\textit{\
we have}%
\begin{equation}
P(0)=1-\rho _{c}-O(\delta \tau ^{1/2}),\text{ }P(1)=\rho _{c}-O(\delta \tau
^{1/2})
\end{equation}%
and for $n\geq 2$%
\begin{equation}
P(n)=\delta \tau \frac{\left( 1-\rho _{c}\right) ^{-1}}{\left(
1-q^{n}\right) \left( 1-q^{n-1}\right) }\left[ 1-\sqrt{\frac{\delta \tau }{%
\left( 1-\rho _{c}\right) (\rho -\rho _{c})}}\right] ^{n}+O(\delta \tau
^{3/2}).
\end{equation}%
For large $n$ the formula for $P(n)$ also yields the exponential decay
\begin{equation}
P(n)\sim \frac{\delta \tau }{\left( 1-\rho _{c}\right) }\exp \left[ -n\sqrt{%
\frac{\delta \tau }{\left( 1-\rho _{c}\right) (\rho -\rho _{c})}}\right] .
\label{cutoffrho>rho_c}
\end{equation}%
However, in this case the cut-off \ depends on $\delta \tau $ diverging as $%
\delta \tau ^{-1/2}$ as $\delta \tau $ goes to zero.

\textit{-- When }$\rho =\rho _{c}$\textit{\ and }$n\ll N$\textit{\ we have}%
\begin{equation}
P(0)=1-\rho _{c}-O(\delta \tau ^{2/3}),\text{ }P(1)=\rho _{c}-O(\delta \tau
^{2/3})  \label{P(n)rho>rho_c}
\end{equation}%
and for $n\geq 2$

\begin{equation}
P(n)\simeq \frac{\delta \tau \left( 1-\rho _{c}\right) ^{-1}}{\left(
1-q^{n}\right) \left( 1-q^{n-1}\right) }\left[ 1-\left( \frac{\delta \tau }{%
\rho _{c}(1-\rho _{c})^{2}}\right) ^{1/3}\right] ^{n}+O\left( \delta \tau
^{4/3}\right) .
\end{equation}%
In this case the cut-off of $P(n)$ at large $n$ is of order of $\delta \tau
^{-1/3}$
\begin{equation}
P(n)\sim \frac{\delta \tau }{\left( 1-\rho _{c}\right) }\exp \left[ -n\left(
\frac{\delta \tau }{\rho _{c}(1-\rho _{c})^{2}}\right) ^{1/3}\right] .
\label{cutoffrho=rho_c}
\end{equation}%
One can see from these formulas that below the critical point, the density
of MP sites at the lattice vanishes proportionally to $\delta \tau $
\begin{equation}
\rho _{MP}(\rho <\rho _{c})\equiv \sum_{n=2}^{M}P(n)\simeq \frac{\delta \tau
}{\rho _{c}-\rho }\frac{\rho ^{2}}{\rho _{c}},
\end{equation}%
while their average occupation number $\left\langle n\right\rangle _{MP}$ is
finite,%
\begin{equation}
\left\langle n\right\rangle _{MP}(\rho <\rho _{c})\equiv \frac{%
\sum_{n=2}^{M}nP(n)}{\sum_{n=2}^{M}P(n)}\simeq \frac{2\rho _{c}-\rho -\rho
\rho _{c}}{\rho _{c}-\rho },
\end{equation}%
both increasing as $\left( \rho _{c}-\rho \right) ^{-1}$ as $\rho $
approaches $\rho _{c}$. Above $\rho _{c\text{ }}$the density of MP sites
vanishes as $\sqrt{\delta \tau }$,
\begin{equation}
\rho _{MP}(\rho >\rho _{c})\simeq \sqrt{\frac{\delta \tau \left( \rho -\rho
_{c}\right) }{(1-\rho _{c})}},
\end{equation}%
which is much slower than in the subcritical region. Their average
occupation number diverges as $\delta \tau ^{-1/2}$
\begin{equation}
\left\langle n\right\rangle _{MP}(\rho >\rho _{c})\simeq \sqrt{\frac{(1-\rho
_{c})\left( \rho -\rho _{c}\right) }{\delta \tau }}.
\end{equation}%
Exactly at the critical point we have
\begin{eqnarray}
\rho _{MP}(\rho  &=&\rho _{c})\simeq \frac{\delta \tau ^{2/3}\rho _{c}^{1/3}%
}{\left( 1-\rho _{c}\right) ^{1/3}}, \\
\left\langle n\right\rangle _{MP}(\rho  &=&\rho _{c})\simeq \left( \frac{%
\rho _{c}}{\delta \tau }\right) ^{1/3}(1-\rho _{c})^{2/3}.
\end{eqnarray}%
One can see that the fraction of particles in MP sites, i.e. $\rho
_{MP}\left\langle n\right\rangle _{MP}$, above the critical point is exactly
equal to $\left( \rho -\rho _{c}\right) $, which preserves the density of
particles in MP sites equal to $\rho _{c}$.

\begin{figure}[tbp]
\unitlength=1mm \makebox(110,45)[cc] {\psfig{file=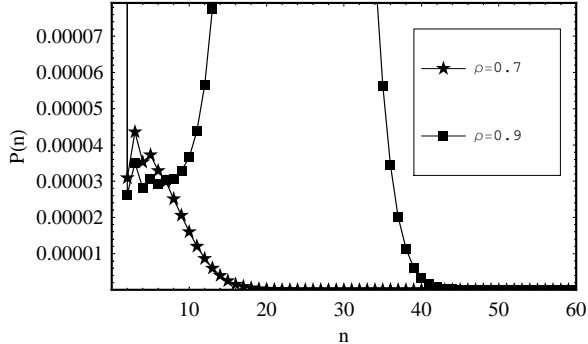,width=80mm}}
\caption{The probability distributions $P(n)$ of the occupation number $n$
for two values of density and $q=0.5,\protect\delta \protect\tau %
=10^{-5},N=100$}
\label{n-distribution}
\end{figure}

The above formulas are derived under the suggestion $n\ll N$. \ For $n$ of
order of $N$ one should take into account the shift of the saddle point in
the numerator of \ Eq.(\ref{P(n)=Z/Z}). The criterion of the applicability
of the saddle point method (\ref{criterion}) suggests
\begin{equation}
\delta \tau ^{1/2}N\gg 1  \label{criterion rho>rho_c}
\end{equation}%
for $\rho >\rho _{c}$ and
\begin{equation}
\delta \tau ^{2/3}N\gg 1  \label{criterion rho=rho_c}
\end{equation}%
for $\rho =\rho _{c}$. Under these conditions the cutoffs of the
distributions (\ref{P(n)rho>rho_c}-\ref{cutoffrho=rho_c}) are much smaller
than $N$, and the probability of MP sites with $n\sim N$ is exponentially
small in $N$. One can address the question what happens beyond the domains (%
\ref{criterion rho>rho_c},\ref{criterion rho=rho_c}). The extreme limiting
case corresponds to the situation of the ASAP on a finite lattice, when $%
\delta \tau $ is so small that one can neglect any quantities of order of $%
\delta \tau ^{2}$. In this case there is a tendency of creation of one MP
site with the occupation number $N(\rho -\rho _{c})$ and the life time that
grows exponentially with the system size $N$ \cite{pph3}. The situation in
the intermediate region should be addressed separately. Direct calculation
of the occupation number distribution \ shows a peak in the middle, which
grows as $\delta \tau $ approaches zero (see Fig (\ref{n-distribution})).
Analysis of the finite lattice behaviour is beyond the goals of present
article.

\section{Summary and discussion.}

We have presented the Bethe ansatz solution of the discrete time ZRP and
ASEP with fully parallel update. We found the eigenfunctions of the equation
for configuration-dependent generating function of the distance travelled by
particles. The eigenfunctions of this equations were looked for in the form
of the Bethe function weighted with the weights of stationary
configurations. We started with the ZRP with arbitrary probabilities of the
hopping of a particle out of a site, which depend on the occupation number
of the site of departure. We inquired which restriction on the probabilities
are imposed by the condition of the Bethe ansatz solvability. As a result we
obtained the two-parametric family of probabilities, the two parameters
being $p,$ the hopping probability of a single particle, and $q$, the
parameter responsible for the dependence of probability on the occupation
number of a site. The model turns out to be very general including as
particular cases $q$-boson asymmetric diffusion model, phase model,
drop-push model and the ASAP. By simple arguments the ZRP can be also
related to the ASEP-like system obeying exclusion interaction. In such
system \ the hopping probabilities depend on the length of the queue of
particles next to a hopping particle. The particular case of the model is
the Nagel-Schreckenberg traffic model with $v_{\max }=1$. We obtained the
Bethe equations for both cases.

The further analysis was devoted to the calculation of generating function
of cumulants of the distance travelled by particles. In the long time limit
it is given by the maximal eigenvalue of the equation discussed. We obtained
the largest eigenvalue in all the phase space of the model. First, the case $%
q=0$ corresponding to the Nagel-Schreckenberg traffic model was considered,
which due to special factorized form of the Bethe equations can be treated
exactly for an arbitrary lattice size. The result is consistent with that
for the continuous-time ASEP and the stationary solution of the
Nagel-Schreckenberg traffic model. For the general $q$ we obtained the
maximal eigenvalue in the scaling limit. It has the universal form specific
for the KPZ universality class. Again, the model dependent constants
obtained reproduce correctly all the known particular cases.

We carried out detailed \ analysis of the limiting case $p\rightarrow 1$.
The phase transition which takes place in this limit was studied. We shown
that below the critical density, the flow of particles consists of \ almost
deterministic synchronous jumps of all particles. Above the critical density
the new phase appears. The finite fraction of all particles gets stuck
immobile at the vanishing fraction of sites. The fluctuations of the
particle current become singular, nonanalytic in $\left( 1-p\right) $.

In terms of the associated ASEP the transition studied is the analytic
continuation of well-known jamming transition in the Nagel-Schreckenberg
model with v$_{\max }=1$, which corresponds to a particular case of our
model. The value of the critical density in that case, $\rho _{c}=1/2$ (or $%
\rho _{c}=1$ for associated ZRP), follows from the particle hole symmetry.
Switching on the interaction between particles breaks this symmetry and as a
result decreases $\rho _{c}$.

From the point of view of the hydrodynamics of the particle flow the reason
of the jamming is the vanishing of the velocity of kinematic waves $%
v_{kin}=\partial \phi /\partial \rho $, which are responsible for the
relaxation of \ inhomogeneities. As was argued in \cite{barmatrip}, the
situation $v_{kin}=0$ leads to the appearance of \ shocks in the ASEP\
interpretation or the MP sites with large occupation numbers\ in ZRP. When $%
\rho >\rho _{c}$ and $p\rightarrow 1$, $v_{kin}$ behaves as follows
\begin{equation*}
v_{kin}=\partial \phi /\partial \rho =\frac{\sqrt{1-p}}{2}\left[ \frac{\rho
_{c}^{2}\left( 1-\rho _{c}\right) }{\left( \rho -\rho _{c}\right) ^{3}}%
\right] ^{1/2},
\end{equation*}%
while in the close vicinity of the \ critical point, $\left\vert \rho -\rho
_{c}\right\vert \sim \left( 1-p\right) ^{1/3}$, the velocity of kinematic
waves vanishes as $\left( 1-p\right) ^{1/3}$. \ Similar transition also was
considered in the bus route model \cite{evans_BRM}. In that case however the
rates $u(n)$ of hopping of a particle out of the site occupied by $n$
particles depend on the external parameter $\lambda $, such that the limits $%
\lambda \rightarrow 0$ and $n\rightarrow \infty $ are not commutative. In
our case the transition owes to taking a limit in only one hopping
probability $p(1)$ and to the discrete parallel update.

We also obtained the canonical partition function of the stationary state of
\ the discrete time ZRP. It is expressed in terms of the $q-$ exponential
series. We apply so called $q-$binomial theorem to evaluate the partition
function in the saddle point approximation. Using the partition function
formalism we obtained the formulas for physical quantities characterizing
the stationary state of the model such as the occupation number distribution
and the average flow of particles, the latter confirming the Bethe ansatz
result. We observed several curious facts, which reveal intimate relation
between the saddle point approximation applied for the partition function
formalism and the thermodynamic limit of the Bethe ansatz solution. We found
that two constants which define the behaviour of the large deviation
function of the particle current can be expressed in terms of the cumulants
of the occupation number of a site and the fugacity of particles.
Furthermore, the analogue of the Helmholtz free energy considered as a
function of arbitrary complex fugacity plays the role of \ a "counter" of
the Bethe roots.\ The equation of state, which relates the density and the
fugacity, coincides with the conical point equation, (\ref{x_c}),which fixes
the position of the Bethe roots contour. All the correspondences found seem
not accidental. \ It was claimed in \cite{da} that the large deviation
function for the particle current should be universal for all models
belonging to the KPZ universality class. On the other hand, once the density
of the Bethe roots is obtained, the derivation of the nonuniversal
parameters is straightforward. It would be natural to expect that the
derivative of the free energy would serve as a generalization of the density
of Bethe roots for the cases where the Bethe ansatz is unapplicable.
Currently we do not have any explanation of the relations found. \ However,
if they do exist, clarifying of their internal structure could give a way
for study of a variety of systems which do not posses the Bethe ansatz
integrability. We leave these questions for further work.

Another possibility of continuation of current study is as follows. In the
particular case, $q=0,$ of our model, which corresponds to the
Nagel-Schreckenberg traffic model with $v_{\max }=1$, the Bethe equations
have special factorized form. Usually, the consequence of such
factorizability is the possibility of time-dependent correlation functions
in the determinant form. This program was realized before for the continuous
time ASEP \cite{schutz} and the ASEP with backward ordered update \cite%
{priezz}. We expect that the parallel update case is also treatable.

\acknowledgments AMP is grateful to V.B. Priezzhev for stimulating
discussion. The work was supported partially by FCT grant
SFR/BPD/11636/2002, projects POCTI/MAT/46176/2002 and POCTI/FAT/46241/2002,
grant of Russian Foundation for Basic Research No.03-01-00780 and DYSONET
project.

\end{document}